\newcommand{\PaperNumber}{462}

\documentclass[10pt, sigconf, letterpaper, nonacm, prologue, table]{acmart}
\usepackage{subfig}
\usepackage{tabularx}
\usepackage{graphicx}
\usepackage{multirow}
\usepackage{tikz}
\usepackage{xcolor}
\usepackage{tablefootnote}
\usepackage{enumitem}
\usepackage{seqsplit}
\usepackage[misc]{ifsym}
\usepackage{pifont}
\usepackage{balance}

\newcommand{\hash}[1]{{\ttfamily\seqsplit{#1}}}

\newcommand{\cmark}{\ding{51}}%
\newcommand{\xmark}{\ding{55}}%
\newcommand{\notcheckmark}{\cmark\makebox[0pt][r]{\kern-1.9ex\raisebox{0.7ex}{\rotatebox[origin=c]{125}{---}}}}

\begin{document}
\title{Ethereum Crypto Wallets under Address Poisoning: How Usable and Secure Are They?}
\author{Shixuan Guan}
\email{sguan6@stevens.edu
}
\affiliation{%
  \institution{Stevens Institute of Technology}
  \city{Hoboken}
  \state{NJ}
  \country{USA}
}

\author{Kai Li}
\email{kli50@stevens.edu}
\affiliation{
  \institution{Stevens Institute of Technology}
  \city{Hoboken}
  \state{NJ}
  \country{USA}
}
\newcommand*\circled[1]{\tikz[baseline=(char.base)]{
            \node[shape=circle,fill,inner sep=1pt] (char) {\textcolor{white}{#1}};}}

\newcommand{\ignore}[1]{}
\begin{abstract}

Blockchain address poisoning is an emerging phishing attack that crafts "similar-looking" transfer records in the victim's transaction history, which aims to deceive victims and lure them into mistakenly transferring funds to the attacker. Recent works have shown that millions of Ethereum users were targeted and lost over 100 million US dollars.

Ethereum crypto wallets, serving users in browsing transaction history and initiating transactions to transfer funds, play a central role in deploying countermeasures to mitigate the address poisoning attack. However, whether they have done so remains an open question. To fill the research void, in this paper, we design experiments to simulate address poisoning attacks and systematically evaluate the usability and security of 53 popular Ethereum crypto wallets. Our evaluation shows that there exist communication failures between 12 wallets and their transaction activity provider, which renders them unable to download the users' transaction history. Besides, our evaluation also shows that 16 wallets pose a high risk to their users due to displaying fake token phishing transfers. Moreover, our further analysis suggests that most wallets rely on transaction activity providers to filter out phishing transfers. However, their phishing detection capability varies. Finally, we found that only three wallets throw an explicit warning message when users attempt to transfer to the phishing address, implying a significant gap within the broader Ethereum crypto wallet community in protecting users from address poisoning attacks.

Overall, our work shows that more efforts are needed by the Ethereum crypto wallet developer community to achieve the highest usability and security standard. Our bug reports have been acknowledged by the developer community, who are currently developing mitigation solutions.

\end{abstract}
\maketitle

\section{Introduction}
The rapid development of the blockchain technology and the flourishing of cryptocurrency markets have brought various scams~\cite{kell2021forsage, bian2021image, xia20covidscams, gao2020tracking, xia21scams, xia20covidscams, vakilinia2022cryptocurrency, xigao2023doublenothing, li2023understanding, li2023towards} and phishing attacks~\cite{chen2020phishing, badawi2020automatic, bartoletti2021cryptocurrency, he2023txphishscope} targeting cryptocurrency holders. Among them, address poisoning is an emerging phishing attack that typically uses "look-alike" addresses to craft phishing transfers to deceive victims, aiming to lure them into mistakenly transferring funds to the attacker. Existing works~\cite{guan24ccs, tsuchiya2025blockchainaddresspoisoning, chen2025dissecting} have reported the \textit{token-based address poisoning attack} on the Ethereum blockchain, where the attackers crafted phishing token transfers to poison victims' ERC-20~\cite{me:erc20} token transfer history. It is shown that attackers crafted three types of phishing token transfers with the "look-alike" address: zero-value, dust-value, and fake token transfers, which caused the victims to lose over 100 million US dollars due to mistakenly transferring funds to the phishing address. More recently, it is reported that attackers also utilized new phishing strategies to attack victims by transferring the native coin, ETH, to poison their transactions history, which we refer to as \textit{ETH-based address poisoning attack}. In this new phishing attack, attackers also crafted three types of phishing transactions: zero-ETH, dust-ETH, and fake-ETH transfers. Such an attack has caused the victim to lose more than 68 million US dollars~\cite{me:68millionloss}. 

\textbf{Research statement:} While the address poisoning attacks have led to such a significant financial loss to victims, it is unknown whether stakeholders in the Ethereum developer community have deployed effective countermeasures to mitigate this threat. Ethereum crypto wallets, which serve as the cornerstone for users to browse transaction history and initiate transactions to transfer funds, thereby play a central role in mitigating the threat. To fill the research gap, in this paper, we choose to measure Ethereum crypto wallets and evaluate their usability and security under the address poisoning attack. Specifically, we aim to answer the following five research questions.

\begin{itemize}[leftmargin=*]
\item\textbf{RQ1:} Are the crypto wallets supporting users to browse their transaction activity history?
\item\textbf{RQ2:} Are the crypto wallets providing good usability to users by displaying their legitimate transfers?
\item\textbf{RQ3:} Are the crypto wallets providing good security to users by hiding or flagging the phishing transfers?
\item\textbf{RQ4:} Who are the wallets’ transaction activity providers? And how do they affect the wallets’ usability and security?
\item\textbf{RQ5:} Are there any preventive countermeasures on the wallets when users transfer to phishing addresses?
\end{itemize}

To answer these questions, we collect 53 popular Ethereum crypto wallets and design experiments to respectively simulate the "token-based" and "ETH-based" address poisoning attacks against them. Specifically, we first generate three Ethereum addresses and then conduct transactions involving both legitimate and phishing transfers of USDT and ETH among them. After our testing transactions are comfirmed on the Ethereum mainnet, we then launch each wallet to import the generated addresses and check if the wallet supports the display of the address's transaction activity. We define three usability levels and five risk levels to respectively quantify the wallet's usability and security under the address poisoning attack. Moreover, we also use Chrome's inspection tool and Wireshark~\cite{me:wireshark} to capture the network traffic of each wallet to identify its transaction activity provider. Finally, on each wallet, we attempt to transfer funds to the phishing address involved in the address poisoning attack to evaluate if there are preventive countermeasures. All our experiments are executed in a contained environment and only affect Ethereum addresses under our control.

\textbf{Research findings:} Our evaluation on the 53 Ethereum crypto wallets reveals several interesting findings. First, we found that 51 out of 53 wallets provide an entry on the wallet's UI to show the users' transaction activity, while only two wallets do not (answers to \textbf{RQ1}). Second, among the 53 wallets, 32 wallets provide good usability and display both legitimate ETH and USDT transfers, and 4 wallets provide fair usability due to only displaying legitimate ETH transfers. In comparison, 17 wallets provide poor usability for not displaying any transfers, including popular wallets such as MetaMask and Crypto.com (answers to \textbf{RQ2}). Third, among the 36 wallets that display legitimate transfers, 6 of them are extremely risky for also displaying all three types of phishing transfers (risk level 4), including the popular Bybit and Core wallets. 10 wallets are highly risky for displaying fake transfers, but not all three types of phishing transfers (risk level 3), including the popular Uniswap and Phantom wallets. In contrast, there are 15 wallets that hide fake transfers. However, they are still risky due to displaying zero-value phishing transfers (risk level 2), including Rabby and Trust wallets. Another 5 wallets are less risky and only display dust-value phishing transfers (risk level 1), such as the Coinbase wallet (answers to \textbf{RQ3}). Moreover, by inspecting the network traffic, our analysis shows that most of the wallets are running their own backend services to feed the transaction activity history, and only 11 wallets are using third-party providers. Among the 9 distinct third-party providers, Etherscan~\cite{me:etherscan} is the most frequently used one. Our further analysis shows that there exist communication failures between 12 wallets and their transaction activity provider, which leads them to be unable to download the transaction history and hence results in their poor usability. In addition, for those that hide one or more phishing transfers, only 10 wallets filter out the phishing transfers solely on the wallet end, while most of them rely on transaction activity providers to filter out phishing transfers. However, the phishing detection capability varies by wallets and transaction activity providers (answers to \textbf{RQ4}). Finally, when attempting to transfer funds to the phishing address, our testing result suggests that only three wallets throw a clear warning message to indicate the risk, which implies a significant security gap in the broader Ethereum crypto wallet community regarding protecting users from address poisoning attacks and safeguarding their funds (answers to \textbf{RQ5}).


In summary, our work makes the following contributions.
\begin{itemize}[leftmargin=*]
\item\textbf{A systematic analysis:} To our best knowledge, we are the first one that systematically evaluates Ethereum crypto wallets under the address poisoning attack by quantifying the wallet's usability and security.
\item\textbf{New understandings:} Our evaluation on 53 Ethereum crypto wallets reveals several interesting findings, including that 17 wallets provide a transaction activity entry but do not display any transfers, 36 wallets are risky for displaying one or more types of phishing transfers. Moreover, we found that there exist communication failures between the wallets and the backend transaction activity providers, and most wallets are relying on the transaction activity provider to filter out phishing transfers, etc.
\item\textbf{New insights:} Our work indicates that no wallet achieves the highest usability and security standard, implying that more efforts are needed by the Ethereum crypto wallet developer community. We also propose what we expect to be achieved in an ideal crypto wallet.
\item\textbf{Bug report:} We have reported the usability bugs and security risks to all the affected crypto wallets. As of this writing, 11 wallets have replied, and 8 wallets have confirmed our report and discussed their mitigation plans.
\end{itemize}

 
\section{Background}
\label{sec:attack}

\begin{figure*}[!htbp]
  \centering
  \centering
  \subfloat[Zero-value transfer.]{%
  \includegraphics[width=0.3\textwidth]{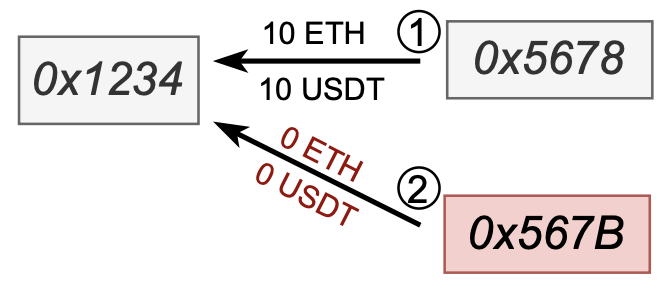}
  \label{fig:zero}}
  \subfloat[Dust-value transfer.]{%
  \includegraphics[width=0.3\textwidth]{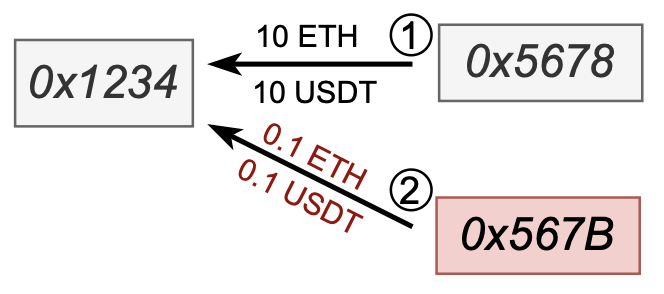}
  \label{fig:dust}}
  \subfloat[Fake token transfer.]{%
  \includegraphics[width=0.3\textwidth]{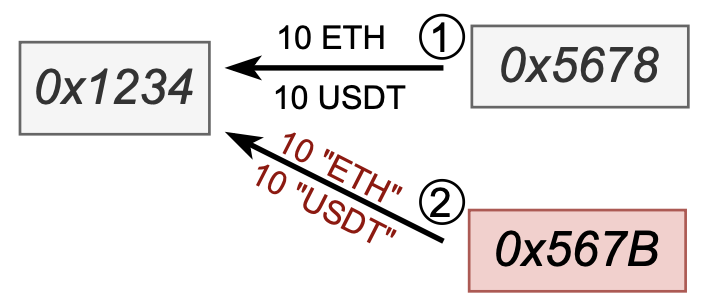}
  \label{fig:fake}}
  \caption{The workflow of Ethereum address poisoning attacks. Upon observing a legitimate transaction between two benign addresses (\protect\circled{1}), the attacker crafts three types of phishing transactions with a similar sender address 0x567B (\protect\circled{2}). The phishing transactions will be added to the receiver's transaction history, who may mistakenly copy the sender's address from the phishing transaction and transfer funds to it.}
  \label{fig:attack_workflow}
\end{figure*}

\subsection{Ethereum Crypto Wallets}
Today, running a full node in the Ethereum blockchain requires a significant amount of computation and storage resources due to the ever-growing ledger size, which makes it unaffordable for average blockchain users. To relieve users' burden of running the full node themselves, various third-party services have been deployed to connect users to the Ethereum blockchain, including RPC services~\cite{me:infura,me:chainstack,me:quicknode,me:alchemy}, blockchain explorers~\cite{me:etherscan,me:oklink,me:blockchaincom}, and crypto wallets. Among them, crypto wallets have become the primary platform utilized by users in managing their crypto assets. In general, crypto wallets can be divided into two categories:
\begin{itemize}[leftmargin=*]
\item \textbf{Custodial wallets} are wallets operated by a third party (such as an exchange or wallet service provider) that directly holds the user's private key. All the user's crypto assets are managed by the third-party, who will sign the transactions on the user's behalf when the payment permission is granted by the user. In this wallet model, the users must trust the third party to securely store their private keys and faithfully sign transactions on their behalf. Examples of custodial wallets provided by exchanges like Binance~\cite{me:binance}, Kraken~\cite{me:kraken}, FreeWallet~\cite{me:freewallet} and BitGo~\cite{me:bitgo}.
\item \textbf{Non-custodial wallets} are wallets that offer users full control over their private keys and crypto assets, without relying on a trusted third party. In this wallet model, the responsibility of securely managing the crypto asset is solely placed on the user. Hence, the security of crypto assets depends on the user’s ability to safeguard their private keys (e.g., safely store and back up the private keys). Examples include browser extension-based wallets~\cite{me:extensionwallet}, hardware wallets~\cite{me:hardwarewallet}, and desktop or mobile apps~\cite{me:appwallet} that allow users to directly control their private keys.
\end{itemize}

In this work, we focus on studying non-custodial wallets as the private key is directly controlled by individual users, which hence makes them the primary target of various scams and phishing attacks. Today, most non-custodial wallets are provided as browser extensions and apps, which can be installed on users' laptops or mobile devices. There are multiple functionalities provided on the wallets. For example, when launching the wallet, users can directly create a new account or import existing accounts from other sources. In addition, on the wallet's UI, users can view the balance of different crypto assets, initiate transactions to transfer the assets, and browse the past transaction activity. In Fig.~\ref{fig:uniswap}, we show such functionalities provided by one of the most popular non-custodial wallets, Uniswap.

\begin{figure}[!htbp]
  \centering
  \includegraphics[width=0.5\textwidth]{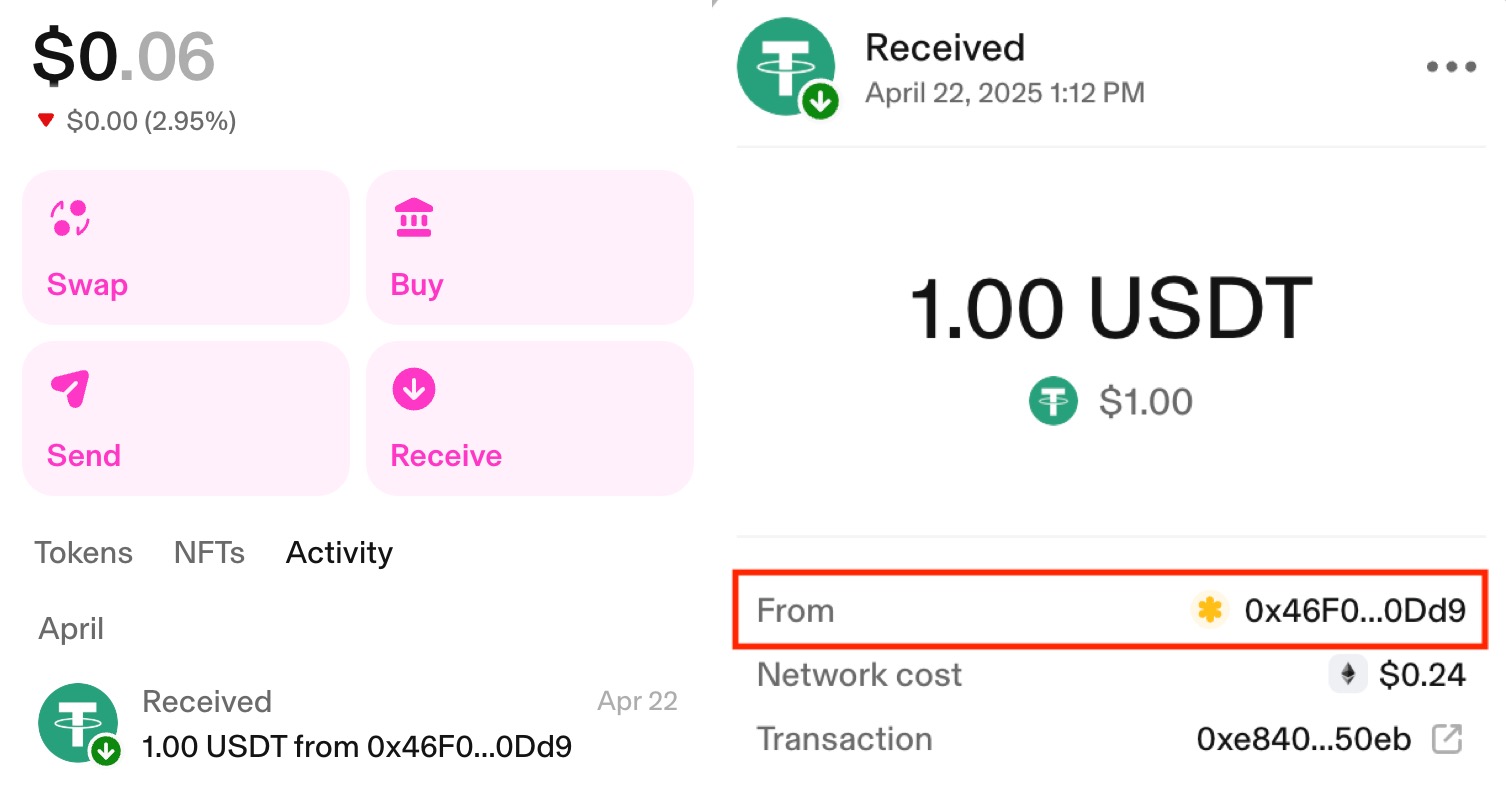}
  \label{fig:uniswap}
  \caption{Address shortening on Uniswap.}%
  \label{fig:uniswap}
\end{figure}
\textbf{Address shortening:} In the Ethereum blockchain, users' accounts are also called Ethereum addresses, which are represented by 40 hexadecimal characters originated from the hash of the user's public key. Due to such a long sequence, it is a common practice for crypto wallets to shorten the address when displaying it on the UI. As shown in Fig.~\ref{fig:uniswap}, Uniswap only shows the first four characters and last four characters of the transaction sender's address, while the transaction recipient's address is not displaying.

\subsection{Ethereum Address Poisoning}
Because users' addresses are shortened on the crypto wallets, they can only rely on the prefix or suffix to differentiate Ethereum addresses, which makes the address poisoning attacks possible. In such an attack, the attacker typically creates a lot of addresses and then monitors transactions recorded on the blockchain. Upon identifying a suitable legitimate transaction where the sender (or receiver) address shares a similar prefix and suffix with the one of the generated addresses ("look-alike" address), the attacker then uses the "look-alike" address to send another transaction (denoted as phishing transaction) to the receiver (or sender) of the legitimate transaction. Since both the legitimate and phishing transactions will be added to the receiver's transaction history, the receiver could be deceived by the phishing transaction and mistakenly transfer assets to the "look-alike" address, resulting in financial loss.

\subsection{Address Poisoning Strategies}
In general, address poisoning attacks on Ethereum can be divided into two categories: token-based address poisoning and ETH-based address poisoning. In the research community, recent works (Guan et al.~\cite{guan24ccs} and Chen et.al~\cite{chen2025dissecting}) have reported that attackers utilized "look-alike" addresses to transfer ERC-20 tokens to victims, which aims to craft a "similar-looking" token transfer record in the victim's token transfer history. We denote such "similar-looking" token transfers sent by attackers as token-based address poisoning attacks. On the social media, a recent post~\cite{me:68millionloss} reported that attackers also utilized "look-alike" addresses to transfer the native coin, ETH, to victims, which aims to craft a "similar-looking" transaction record in the victims' transaction history. We thereby denote the "similar-looking" ETH transfer as ETH-based address poisoning attacks. Fig.~\ref{fig:attack_workflow} shows the typical workflow of such two types of address poisoning attacks. In both attacks, the attacker can craft three types of phishing transfers to deceive victims: zero-value transfer, dust-value transfer, and fake-token transfer.

\textbf{Zero-value transfer:} Upon finding a legitimate transaction received by the victim in step \protect\circled{1}, the attacker utilizes a phishing address \textit{0x567B} to send zero amount of valuable token or ETH to the victim's address (\protect\circled{2}). After that, the victim \textit{0x1234} will observe two transactions displayed on the crypto wallet: the legitimate transfer from \textit{0x5678}, and the zero-value transfer from \textit{0x567B}. Since such two addresses look highly similar, when the victim decides to transfer crypto assets back to the legitimate address \textit{0x5678}, she/he may mistakenly copy the phishing address \textit{0x567B} and transfer funds to it, resulting in financial loss.

\textbf{Dust-value transfer:} This phishing transfer works similarly to the above and only differs in step \protect\circled{2}. That is, instead of transferring zero amount, the attacker uses the phishing address to transfer a tiny amount of valuable token or ETH to the victim. Likewise, such a phishing transfer will be displayed on the victim's crypto wallets, which could mislead them to transfer funds to the phishing address, resulting in financial loss. Compared to the zero-value transfer, while dust-value transfer incurs a higher cost, it could increase the attacker's success rate as the transferred amount in the phishing transaction can be adjusted to look more similar to the transferred amount in the legitimate transaction.

\textbf{Fake-token transfer:} Since the transferred amount in the above two phishing transfers cannot match exactly with the amount of the legitimate transaction, an economic solution adopted by the attacker is to deploy fake ERC20 tokens or "ETH" tokens and craft phishing transfers that have the same transferred amount. After observing a legitimate transaction, the attacker then uses the phishing address \textit{0x567B} to transfer "fake" amount to the victim address \textit{0x1234}. Such a phishing transfer will also be displayed on the victim's crypto wallets and potentially lead them to transfer funds to the attacker. 
 
\section{Ethereum Wallet Measurement}
\label{sec:wallet}
As described in Sec.~\ref{sec:attack}, both the token-based and ETH-based address poisoning attacks aim to poison victims' transaction history. Given the fact that crypto wallets are the primary platform utilized by cryptocurrency users in browsing transaction history and transferring funds, it is thus imperative for them to deploy countermeasures to mitigate the threat of address poisoning attacks. However, to the best of our knowledge, it remains an open question whether the crypto wallets have done so to protect their users. In this paper, we aim to answer this question by systematically testing popular Ethereum crypto wallets and evaluating their usability and security under the address poisoning attack.


\subsection{Methodology}
\begin{figure}[!ht]
   \centering
   \includegraphics[width=.38\textwidth]{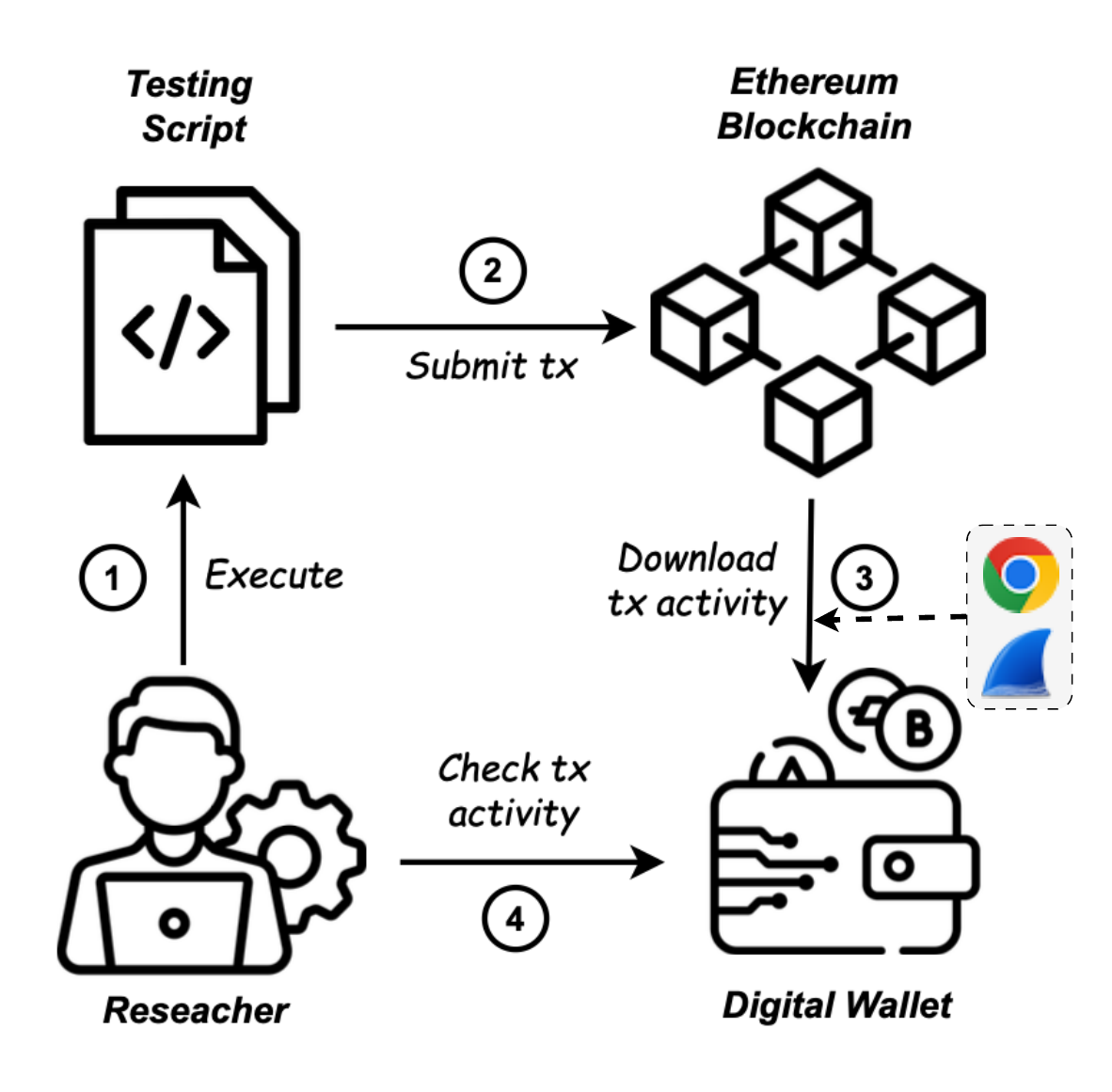}
   \caption{Our wallet evaluation methodology.}%
   \label{fig:wallet_test}
\end{figure}

To evaluate the usability and security of Ethereum crypto wallets, we choose to download them and simulate the actual address poisoning attack against a victim address under our control. Fig.~\ref{fig:wallet_test} shows our evaluation methodology. After installing the crypto wallet apps, we execute a testing script to send both legitimate and phishing transactions to the victim address. After the transactions are included in the Ethereum mainnet, we then launch each wallet and import the victim address to download the transaction history. Meanwhile, we use Chrome's inspection tool and Wireshark~\cite{me:wireshark} to capture the wallet's network traffic. Finally, we enter the "transaction activity" entry on the wallet to check if legitimate and phishing transactions are displayed. 

\subsection{Measurement Process}
\textbf{Preparation:} We begin by leveraging the public crypto wallet listing and ranking platforms CryptoSlate~\cite{me:cryptoslate} and Alchemy~\cite{me:alchemy} to collect Ethereum crypto wallets. In total, we download 53 Ethereum crypto wallets and install them on our laptop. Among them, 51 wallets are Chrome Extensions (CE), and two wallets are Desktop apps. After that, we develop a testing script to simulate the address poisoning attack with the following three externally-owned addresses (EOA).
\begin{itemize}
\item \textbf{V}: victim address that receives funds;
\item \textbf{B}: Benign address that transfers funds to \textbf{V};
\item \textbf{P}: phishing address that looks like \textbf{B}, which will launch the address poisoning attack against \textbf{V};
\end{itemize}
We obtain Address \textbf{V} (\hash{0x71aF257EF2fA722694E1621B6f1D968c28Dd7A95}) by directly creating an account on the MetaMask wallet. To create the two similar addresses (\textbf{B} and \textbf{P}), we implement a Python script to invoke the \texttt{eth\_createAccount} library function to generate addresses. After executing the function more than 200K times, we obtain a suitable \textbf{B} (\hash{0x46F0196EdBb29Bd3715E7F556c8633efDe1D0Dd9}) and \textbf{P} (\hash{0x46F0042749ad2383471639b57833cd80bf1f0Dd9}), which share the same four characters at the beginning and ending segments. After that, we also need to identify the following three token contract addresses on the Ethereum mainnet.

\begin{itemize}
    \item \textbf{Legit USDT contract}: to emit legit-USDT, zero-USDT, and dust-USDT transfers;
    \item \textbf{Fake USDT contract}: to emit fake-USDT transfers;
    \item \textbf{Fake "ETH" contract}: to emit fake-ETH transfers;
\end{itemize}
The legitimate USDT contract is well-known at address \hash{0xdAC17F958D2ee523a2206206994597C13D831ec7}. For the fake USDT and ETH contracts, although we can deploy them on the Ethereum mainnet ourselves, we chose not to. Instead, we leverage the existing fake tokens already deployed by the attackers, thereby avoiding further pollution of the Ethereum mainnet. From the thousands of fake tokens flagged on Etherscan~\cite{me:etherscan}, we test each flagged token contract locally until we find one that allows us to send transactions to execute the \texttt{transferFrom} function with an arbitrary sender, recipient, and amount. Through testing, we found a suitable fake USDT token deployed at \hash{0xF27d16****eA2A85} and a fake "ETH" token deployed at \hash{0x046674****fC7445}.

\textbf{Send testing transactions:} After installing the crypto wallets and obtaining the addresses, we import the private key of \textbf{V} into each wallet and then execute the testing script to send transactions to transfer ETH and USDT from \textbf{B} and \textbf{P} to \textbf{V}. Two sets of transactions are sent to respectively simulate the ETH-based and token-based address poisoning attacks, each containing five transactions as shown in Table~\ref{tab:eth_transactions}. TX1 simulates a legitimate ETH transfer from \textbf{B} to \textbf{V}, while TX2 - TX4 simulate zero-ETH, dust-ETH, and fake-ETH transfers from \textbf{P} to \textbf{V}. TX5 simulates a special fake-ETH transfer with zero amount. Similarly, TX6 simulates a legitimate USDT transfer, and TX7 - TX9 simulate zero-USDT, dust-USDT, and fake-USDT transfers. TX10 simulates the special fake-USDT transfer with zero amount.
\begin{table}[!htbp]
\caption{Transactions simulating the ETH-based and token-based address poisoning attacks.}
\label{tab:eth_transactions}
\centering
\resizebox{\columnwidth}{!}{%
\begin{tabular}{|c|c|c|c|c|l|}
\hline
\textbf{TX} & \textbf{From} & \textbf{To} & \textbf{Amount} & \textbf{Transfer} & \multicolumn{1}{c|}{\textbf{Hash}} \\ \hline
1           & B             & V           & 0.001           & legit ETH         & 0x06ae70...db8dec                  \\ \hline
2           & P             & V           & 0               & zero-ETH          & 0x453460...92dc5c                  \\ \hline
3           & P             & V           & 0.00001         & dust-ETH          & 0xf91b8a...97de23                  \\ \hline
4           & P             & V           & 0.001           & fake-ETH          & 0xb0dc19...fe05cd                  \\ \hline
5           & P             & V           & 0               & fake-ETH          & 0x61166f...551128                  \\ \hline
6           & B             & V           & 10              & legit USDT        & 0xb4041d...5aafe7                  \\ \hline
7           & P             & V           & 0               & zero-USDT         & 0x7c076b...c646b9                  \\ \hline
8           & P             & V           & 0.01            & dust-USDT         & 0x93a38c...4b6892                  \\ \hline
9           & P             & V           & 10              & fake-USDT         & 0xe3dc33...9fefa0                  \\ \hline
10          & P             & V           & 0               & fake-USDT         & 0x5a7042...a53672                  \\ \hline
\end{tabular}
}
\end{table}

\begin{figure*}[!htbp]
  \centering
    \subfloat[One entry per coin (Safepal)]{%
  \includegraphics[width=0.293\textwidth]{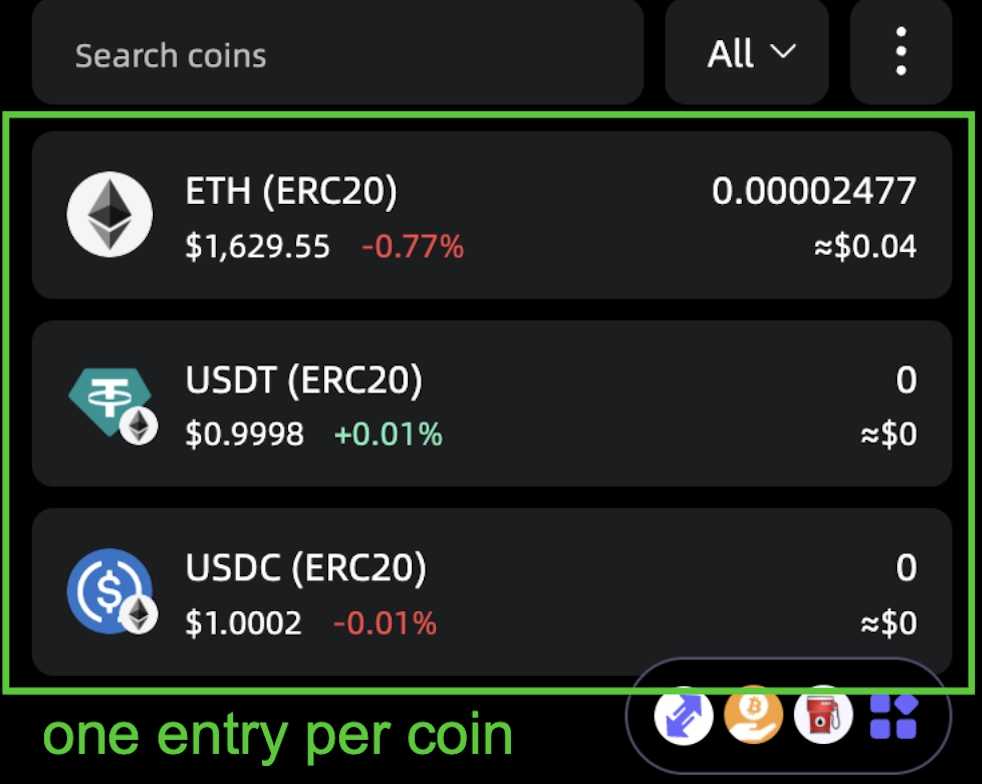}
  \label{fig:onepercoin}
    }%
   \subfloat[One entry for all (Ctrl)]{%
  \includegraphics[width=0.300\textwidth]{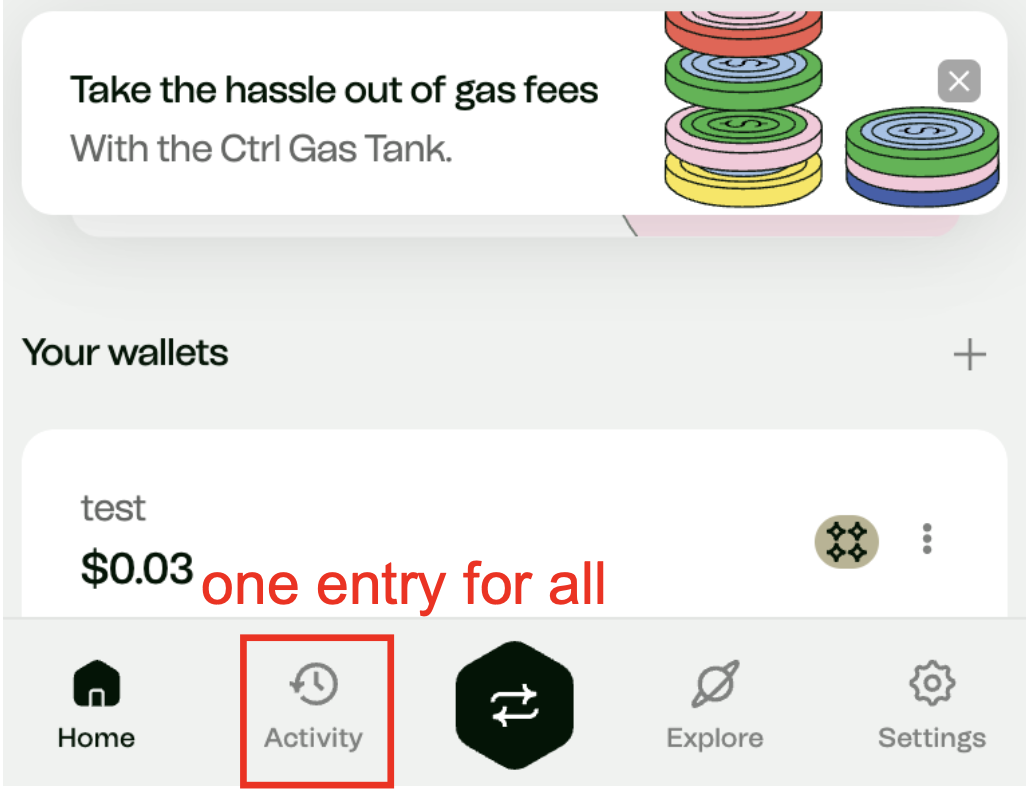}
  \label{fig:oneforall}}
   \subfloat[Hybrid (MetaMask)]{%
  \includegraphics[width=0.336\textwidth]{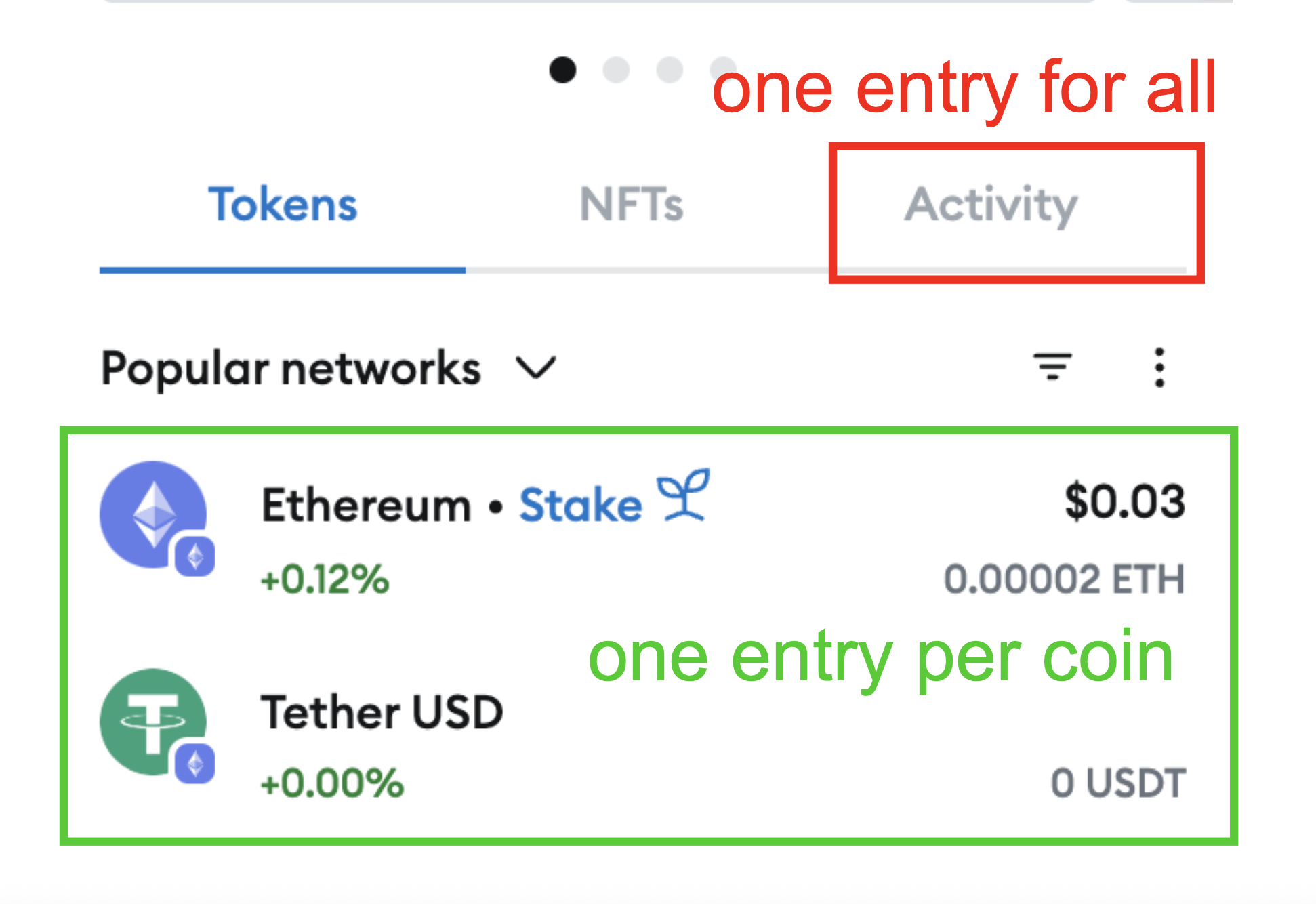}
  \label{fig:hybrid}}
  \caption{Three designs of transaction activity entry.}%
  \label{fig:entry}
\end{figure*}
\textbf{Launch the wallet app:} After our testing transactions are included on the Ethereum mainnet, we launch each crypto wallet on our laptop and enter the "transaction activity" tab to observe if the testing transactions are displayed. We check the following three conditions to assess both the wallet's usability and security: \textbf{C1}: the legitimate transfers are displayed; \textbf{C2}: the phishing transfers are displayed; \textbf{C3}: the addresses in each transfer are shortened.

\section{Wallet Measurement Results}
This section details our measurement results of 53 Ethereum crypto wallets, including the designs of transaction activity entry, display of legitimate and phishing transfers, transaction activity provider, and transaction warning features.  

\begin{table}[]
\caption{Designs of transaction activity entry.}
\label{tab:entry}
\resizebox{\columnwidth}{!}{%
\begin{tabular}{|c|c|}
\hline
\textbf{Tx Activity} & \textbf{Wallets}                                                                                                                                                  \\ \hline
No entry                     & Frame, Cypher                                                                                                                                                  \\ \hline
One entry per coin                     & \begin{tabular}[c]{@{}c@{}} Coin98, Klever, SafePal, \\ FoxWallet, Nabox, Tomato \end{tabular} \\ \hline
One entry for all                     & \begin{tabular}[c]{@{}c@{}}Ctrl, Enkrypt, Frontier, \\ Zeal, Phantom, Rabby, Rainbow, \\ Sender, SubWallet, Uniswap, Virgo, \\ WigWam, Zerion, Nest, Quantum \\ Nightly, Nufi, JustLiquidity, Safnect  \end{tabular}                                  \\ \hline
Hybrid                                  & \begin{tabular}[c]{@{}c@{}}Coinbase, Bybit, Core, \\ Backpack, Gate, Hana, Onekey, \\ Taho, Exodus, OKX, TokenPocket, \\ Trust, Pontem, Bitget, MetaMask, \\ Aurox, Crypto.com, Keplr, Infinity, \\ Nufinetes, Superhero, Wombat, Zapit\end{tabular} \\ \hline
\end{tabular}
}
\end{table}

\subsection{Design of Transaction Activity Entry}
By launching the 53 wallets to check the transaction activity, our first observation is that several crypto wallets do not provide a UI entry to display users' transaction activity. In addition, crypto wallets that provide this functionality have different UI designs for displaying transaction activity. In Table~\ref{tab:entry}, we summarize our observations.

Among the 53 crypto wallets, Frame and Cypher are the only two wallets that do not provide an entry to show users' transaction activity on the UI. For the remaining 51 wallets providing such functionality, their transaction activity entries generally have three types of designs. The first type is "one entry per coin", where the wallets separate the transaction activity by the type of crypto assets. As shown in Fig.~\ref{fig:onepercoin}, in such a design, the native coin "ETH" has its own entry for all "ETH" transfers, and each popular ERC-20 tokens, such as USDT and USDC, also have its own entry to show the corresponding transaction activity. Users can also manually create a new entry to add other tokens. Our observation shows that 6 wallets, including Klever, Safepal, and FoxWallet, etc, employed such a design. The second type is "one entry for all", where the wallets simply merge all transfers into one entry, regardless of the type of crypto assets, as shown in Fig.~\ref{fig:oneforall}. In this design, users cannot manually create an entry for a specific token. Our result indicates that 19 wallets, such as Ctrl, Phantom, Uniswap, Zerion, Nightly, etc, adopted such a design. The last type is a hybrid design that supports both "one entry per coin" and "one entry for all". It thus depends on the users' preference to select an entry to browse the transaction activity. Fig.~\ref{fig:hybrid} shows an example of such a hybrid design. Our observation suggests that 23 wallets have employed a hybrid design, including Coinbase, Backpack, Trust, MetaMask, Crypto.com, etc.

\subsection{Display of Legitimate Transfers}
We assess each wallet's usability by checking condition \textbf{C1}: display of legitimate transfers. Since there are two types of legitimate transfers (ETH and USDT) that could be displayed in the transaction activity entries, we define the following criteria to quantify the usability. The higher the level is, the better usability the wallet can provide. For wallets employing a hybrid design of the transaction activity entry, we enter all possible entries to assign the usability level. 
\begin{itemize}
\item \textbf{Level 2}: display legitimate ETH and USDT transfers;
\item \textbf{Level 1}: display only one legitimate transfer;
\item \textbf{Level 0}: display none of the legitimate transfers;
\end{itemize}

\textbf{Findings:} Our measurement result shows that 17 wallets do not display any transfers, including Frame, 1chainAi, Crypto.com, MetaMask, Nightly, SubWallet, as listed in Table~\ref{tab:wallet_results_zero}. Due to this, their usability level is 0. It is also unknown whether they would shorten the user's address. All the remaining 36 wallets display the legitimate transfers, as shown in Table~\ref{tab:wallet_results}. Among them, 4 wallets provide usability at level 1 due to only displaying the legitimate ETH transfer, such as Enkrypt, FoxWallet, Klever, and Pontem. The other 32 wallets provide usability at level 2 for displaying both legitimate ETH and USDT transfers. It is also interesting to see that no wallets would only display the legitimate USDT transfer without showing the legitimate ETH transfer, which implies that displaying ETH transfers has a higher priority than token transfers. This can be explained by the fact that ETH is the fiat currency on Ethereum, and ETH transfers are more frequent than token transfers. Besides, when transferring tokens, users must have ETH in their accounts to pay the transaction fee. Due to these reasons, supporting ETH transfer history has a higher priority than token transfer history in the design of Ethereum crypto wallets.

\begin{table}[!htbp]
\caption{17 wallets that do not display any transfers. "CE" indicates the wallet is a Chrome Extension.}
\label{tab:wallet_results_zero}
\resizebox{\columnwidth}{!}{
\begin{tabular}{|c|c|c|c|}
\hline
\textbf{Wallet}                                                                                                                                                                    & \textbf{Type} & \textbf{\begin{tabular}[c]{@{}c@{}}Usability\\ Level\end{tabular}} & \textbf{\begin{tabular}[c]{@{}c@{}}Risk\\ Level\end{tabular}} \\ \hline
Frame                                                                                                                                                                              & Desktop           & 0                                                                  & 0                                                             \\ \hline
\begin{tabular}[c]{@{}c@{}}1chainAi, Aurox, Arcana, \\ Crypto.com, Cypher, Fluvi, \\ Keplr, JustLiquidity, MetaMask, \\ Nabox, Nightly, Nufinetes,\\ Quantum, SubWallet,Virgo, Zapit\end{tabular} & CE            & 0                                                                  & 0                                                             \\ \hline
\end{tabular}%
}
\end{table}

\begin{figure*}[!htbp]
  \centering
    \subfloat[Bybit]{%
  \includegraphics[width=0.21\textwidth]{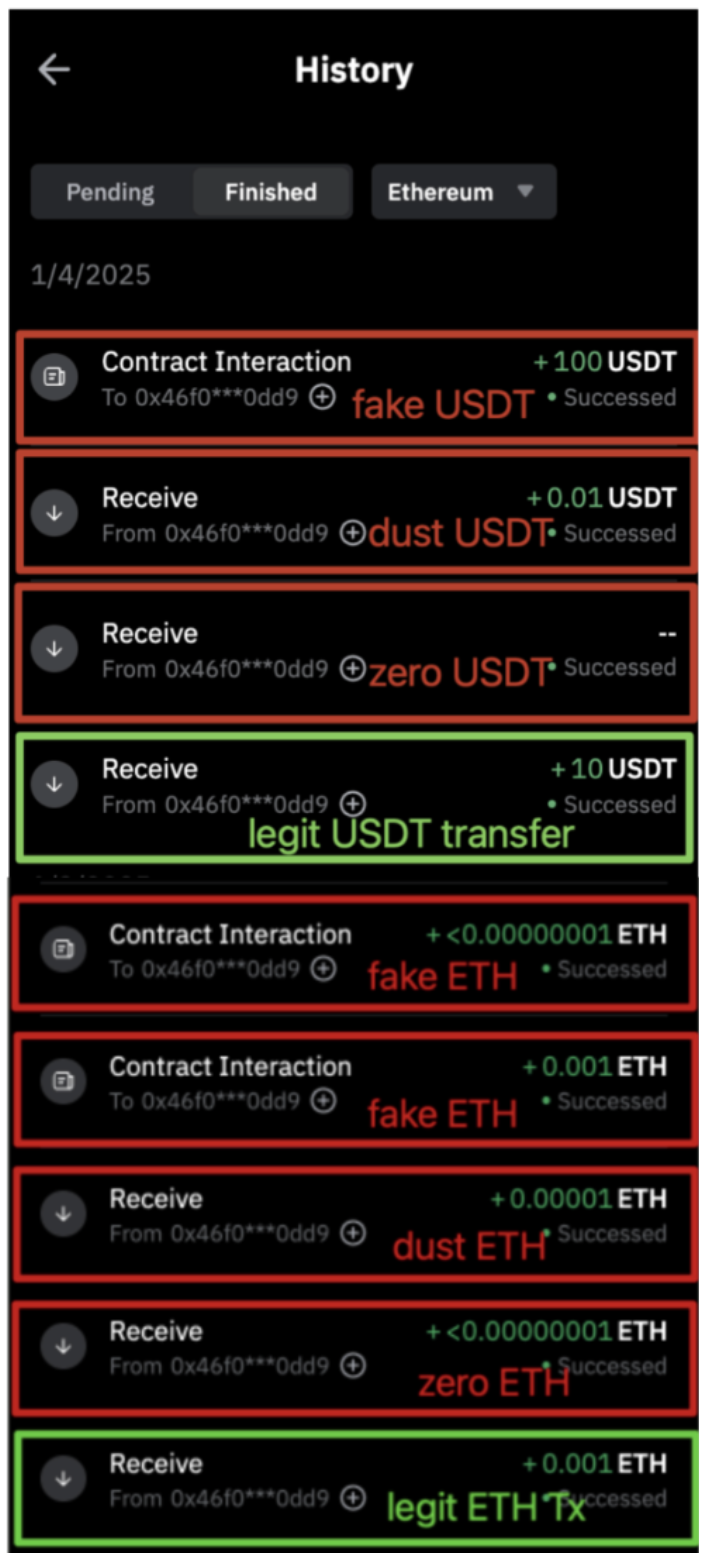}
  \label{fig:bybit}
    }%
   \subfloat[Core]{%
  \includegraphics[width=0.345\textwidth]{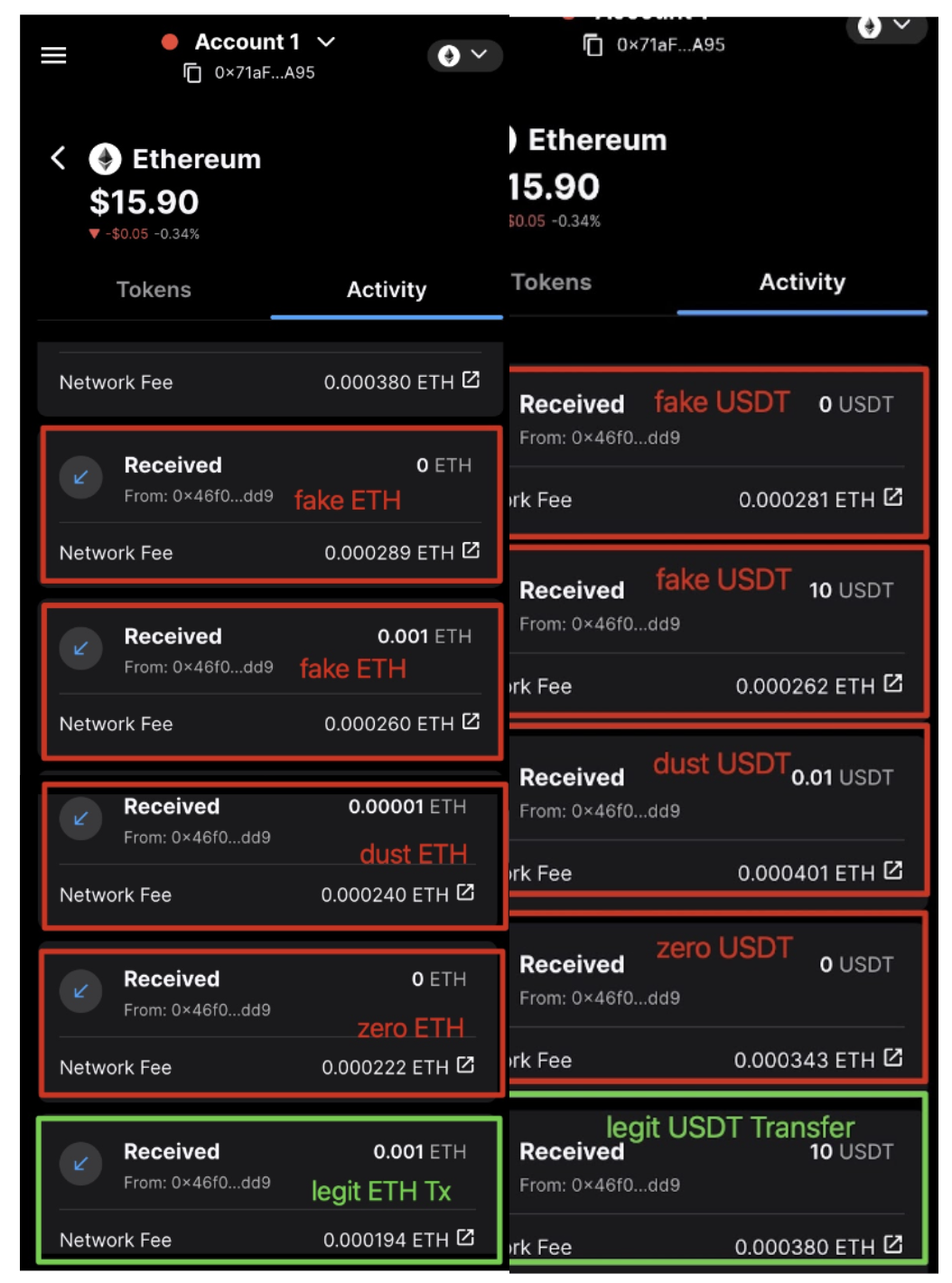}
  \label{fig:core}}
   \subfloat[Ctrl]{%
  \includegraphics[width=0.235\textwidth]{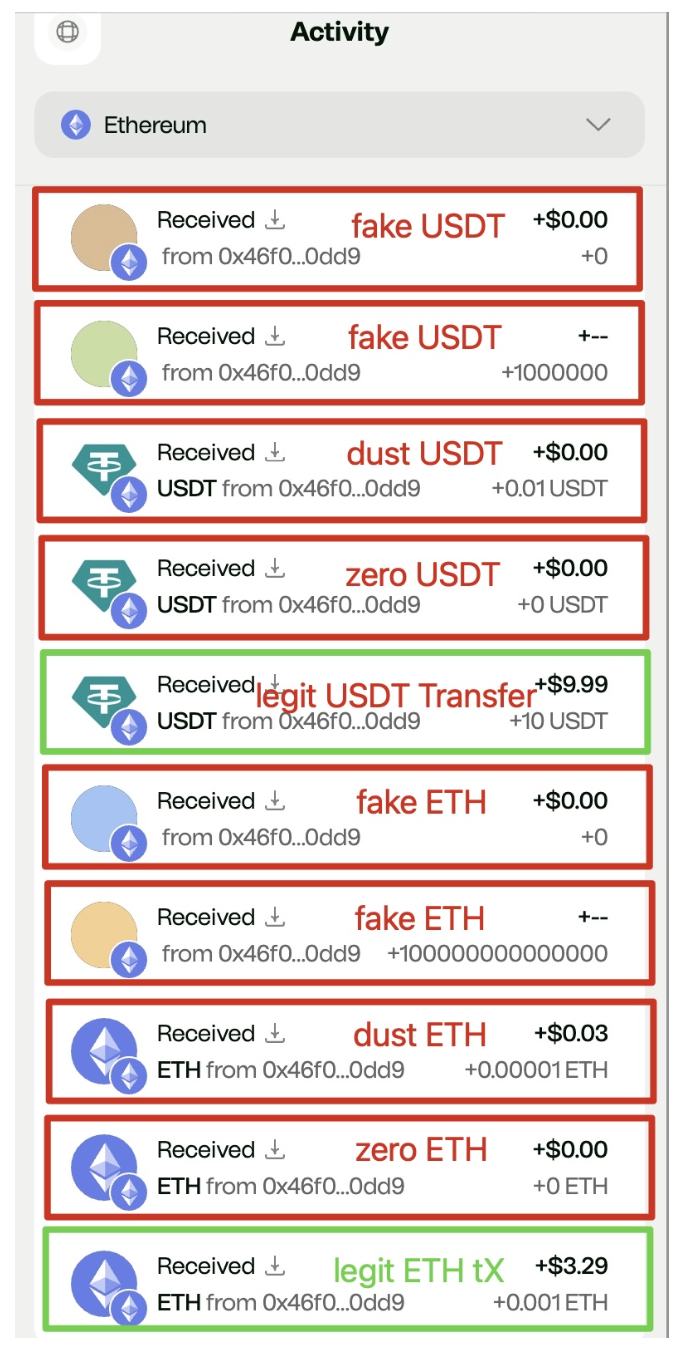}
  \label{fig:ctrl}}
   \subfloat[Frontier]{%
  \includegraphics[width=0.23\textwidth]{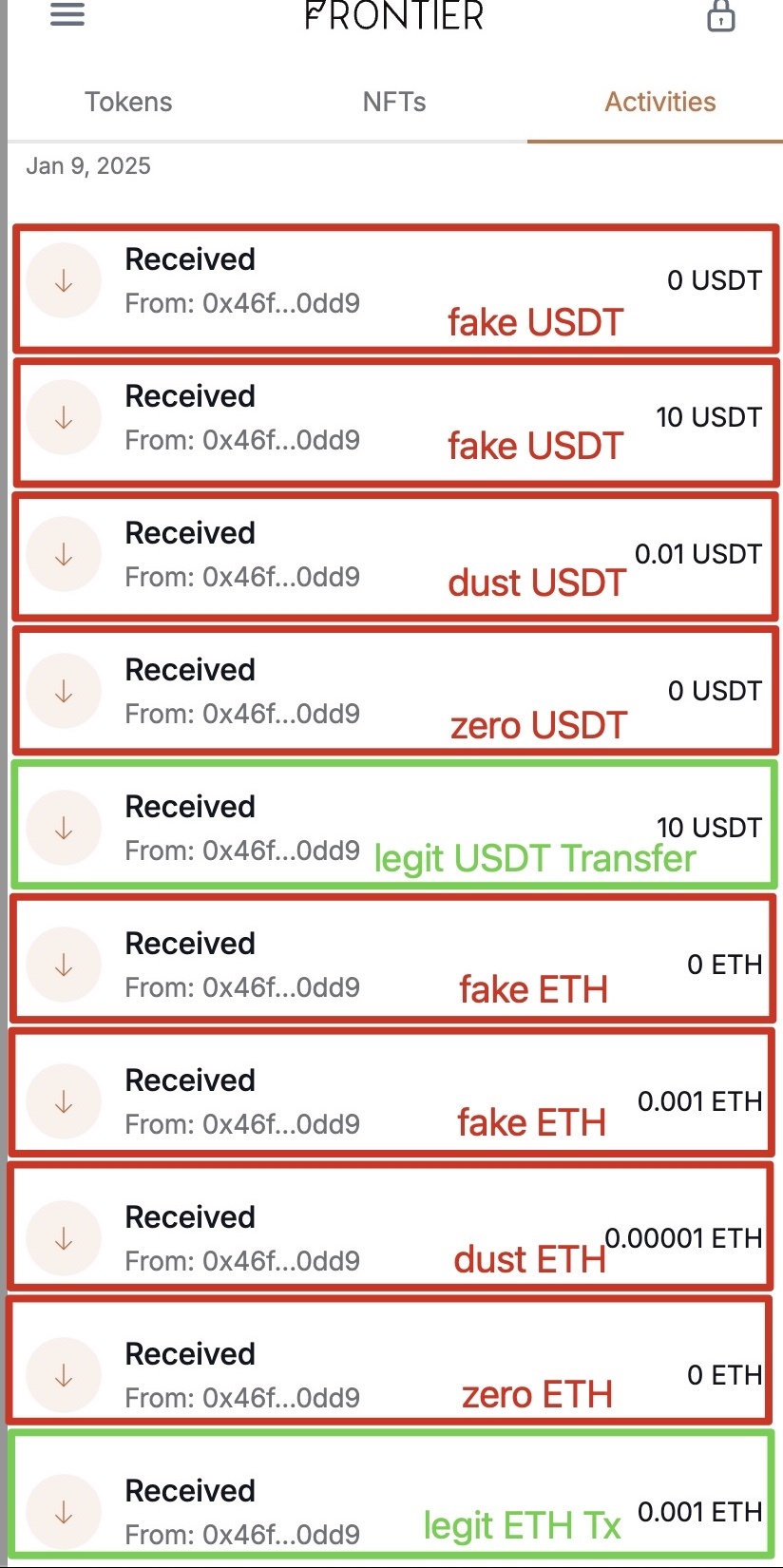}
  \label{fig:frontier}}
  \caption{Transfers displayed by four extremely risky wallets (risk level=4).}%
  \label{fig:level4}
\end{figure*}

\begin{table*}[!htbp]
\caption{Measurement results of 36 Ethereum crypto wallets that display transfers. {\cmark}  indicates the transfer is displayed. {\xmark}  indicates the transfer is not displayed. \notcheckmark indicates the fake transfer is conditionally displayed.}
\label{tab:wallet_results}
\centering
\resizebox{\textwidth}{!}{
\begin{tabular}{|c|c|cccc|cccc|c|c|c|}
\hline
\multirow{2}{*}{\textbf{Wallet}} & \multirow{2}{*}{\textbf{Type}} & \multicolumn{4}{c|}{\textbf{ETH Transfers}}                                                                                                                         & \multicolumn{4}{c|}{\textbf{USDT Transfers}}                                                                                                                        & \multirow{2}{*}{\textbf{\begin{tabular}[c]{@{}c@{}}Address \\ Shorten\end{tabular}}} & \multirow{2}{*}{\textbf{\begin{tabular}[c]{@{}c@{}}Usability\\ Level\end{tabular}}} & \multirow{2}{*}{\textbf{\begin{tabular}[c]{@{}c@{}}Risk\\ Level\end{tabular}}} \\ \cline{3-10}
                                 &                                & \multicolumn{1}{c|}{\textbf{Legit}}        & \multicolumn{1}{c|}{\textbf{Zero}}         & \multicolumn{1}{c|}{\textbf{Dust}}         & \textbf{Fake}                & \multicolumn{1}{c|}{\textbf{Legit}}        & \multicolumn{1}{c|}{\textbf{Zero}}         & \multicolumn{1}{c|}{\textbf{Dust}}         & \textbf{Fake}                &                                                                                      &                                                                                     &                                                                                \\ \hline
\rowcolor[HTML]{A9A7A7}
Bybit                            & CE                             & \multicolumn{1}{c|}{\cmark} & \multicolumn{1}{c|}{\cmark} & \multicolumn{1}{c|}{\cmark} & \cmark        & \multicolumn{1}{c|}{\cmark} & \multicolumn{1}{c|}{\cmark} & \multicolumn{1}{c|}{\cmark} & \cmark        & \cmark                                                                & 2                                                                                   & 4                                                                              \\ \hline
\rowcolor[HTML]{A9A7A7}
Core                             & CE                             & \multicolumn{1}{c|}{\cmark} & \multicolumn{1}{c|}{\cmark} & \multicolumn{1}{c|}{\cmark} & \cmark        & \multicolumn{1}{c|}{\cmark} & \multicolumn{1}{c|}{\cmark} & \multicolumn{1}{c|}{\cmark} & \cmark        & \cmark                                                                & 2                                                                                   & 4                                                                              \\ \hline
\rowcolor[HTML]{A9A7A7}

Ctrl                             & CE                             & \multicolumn{1}{c|}{\cmark} & \multicolumn{1}{c|}{\cmark} & \multicolumn{1}{c|}{\cmark} & \cmark        & \multicolumn{1}{c|}{\cmark} & \multicolumn{1}{c|}{\cmark} & \multicolumn{1}{c|}{\cmark} & \cmark        & \cmark                                                                & 2                                                                                   & 4                                                                              \\ \hline
\rowcolor[HTML]{A9A7A7}
Frontier                         & CE                             & \multicolumn{1}{c|}{\cmark} & \multicolumn{1}{c|}{\cmark} & \multicolumn{1}{c|}{\cmark} & \cmark        & \multicolumn{1}{c|}{\cmark} & \multicolumn{1}{c|}{\cmark} & \multicolumn{1}{c|}{\cmark} & \cmark        & \cmark                                                                & 2                                                                                   & 4                                                                              \\ \hline
\rowcolor[HTML]{A9A7A7}
Safnect   & CE & \multicolumn{1}{c|}{\cmark} & \multicolumn{1}{c|}{\cmark} & \multicolumn{1}{c|}{\cmark} & \cmark & \multicolumn{1}{c|}{\cmark} & \multicolumn{1}{c|}{\cmark} & \multicolumn{1}{c|}{\cmark} & \cmark & \cmark & 2 & 4 \\ \hline
\rowcolor[HTML]{A9A7A7}
Superhero & CE & \multicolumn{1}{c|}{\cmark} & \multicolumn{1}{c|}{\cmark} & \multicolumn{1}{c|}{\cmark} & \cmark & \multicolumn{1}{c|}{\cmark} & \multicolumn{1}{c|}{\cmark} & \multicolumn{1}{c|}{\cmark} & \cmark & \cmark & 2 & 4 \\ \hline
\rowcolor[HTML]{B6B6B6}

Backpack                         & CE                             & \multicolumn{1}{c|}{\cmark} & \multicolumn{1}{c|}{\xmark} & \multicolumn{1}{c|}{\cmark} & \cmark        & \multicolumn{1}{c|}{\cmark} & \multicolumn{1}{c|}{\xmark} & \multicolumn{1}{c|}{\cmark} & \cmark        & \cmark                                                                & 2                                                                                   & 3                                                                              \\ \hline
\rowcolor[HTML]{B6B6B6}
Gate                             & CE                             & \multicolumn{1}{c|}{\cmark} & \multicolumn{1}{c|}{\xmark} & \multicolumn{1}{c|}{\cmark} & \cmark        & \multicolumn{1}{c|}{\cmark} & \multicolumn{1}{c|}{\xmark} & \multicolumn{1}{c|}{\cmark} & \cmark        & \cmark                                                                & 2                                                                                   & 3                                                                              \\ \hline
\rowcolor[HTML]{B6B6B6}
Hana                             & CE                             & \multicolumn{1}{c|}{\cmark} & \multicolumn{1}{c|}{\xmark} & \multicolumn{1}{c|}{\cmark} & \cmark        & \multicolumn{1}{c|}{\cmark} & \multicolumn{1}{c|}{\cmark} & \multicolumn{1}{c|}{\cmark} & \cmark        & \cmark                                                                & 2                                                                                   & 3                                                                              \\ \hline
\rowcolor[HTML]{B6B6B6}
OneKey                           & CE                             & \multicolumn{1}{c|}{\cmark} & \multicolumn{1}{c|}{\cmark} & \multicolumn{1}{c|}{\cmark} & \cmark        & \multicolumn{1}{c|}{\cmark} & \multicolumn{1}{c|}{\cmark} & \multicolumn{1}{c|}{\cmark} & \xmark        & \cmark                                                                & 2                                                                                   & 3                                                                              \\ \hline
\rowcolor[HTML]{B6B6B6}
Nest                           & CE                             & \multicolumn{1}{c|}{\cmark} & \multicolumn{1}{c|}{\cmark} & \multicolumn{1}{c|}{\cmark} & \xmark        & \multicolumn{1}{c|}{\cmark} & \multicolumn{1}{c|}{\xmark} & \multicolumn{1}{c|}{\cmark} & \cmark        & \cmark                                                                & 2                                                                                   & 3                                                                              \\ \hline

\rowcolor[HTML]{B6B6B6}

Nufi      & CE & \multicolumn{1}{c|}{\cmark} & \multicolumn{1}{c|}{\xmark} & \multicolumn{1}{c|}{\cmark} & \cmark & \multicolumn{1}{c|}{\cmark} & \multicolumn{1}{c|}{\xmark} & \multicolumn{1}{c|}{\cmark} & \cmark & \cmark & 2 & 3 \\ \hline
\rowcolor[HTML]{B6B6B6}

Phantom                          & CE                             & \multicolumn{1}{c|}{\cmark} & \multicolumn{1}{c|}{\cmark} & \multicolumn{1}{c|}{\cmark} & \cmark        & \multicolumn{1}{c|}{\cmark} & \multicolumn{1}{c|}{\cmark} & \multicolumn{1}{c|}{\cmark} & \xmark        & \cmark                                                                & 2                                                                                   & 3                                                                              \\ \hline
\rowcolor[HTML]{B6B6B6}
Taho                             & CE                             & \multicolumn{1}{c|}{\cmark} & \multicolumn{1}{c|}{\cmark} & \multicolumn{1}{c|}{\cmark} & \cmark        & \multicolumn{1}{c|}{\cmark} & \multicolumn{1}{c|}{\xmark} & \multicolumn{1}{c|}{\cmark} & \cmark        & \cmark                                                                & 2                                                                                   & 3                                                                              \\ \hline
\rowcolor[HTML]{B6B6B6}

Uniswap                          & CE                             & \multicolumn{1}{c|}{\cmark} & \multicolumn{1}{c|}{\cmark} & \multicolumn{1}{c|}{\cmark} & \cmark        & \multicolumn{1}{c|}{\cmark} & \multicolumn{1}{c|}{\xmark} & \multicolumn{1}{c|}{\cmark} & \cmark        & \cmark                                                                & 2                                                                                   & 3                                                                              \\ \hline
\rowcolor[HTML]{B6B6B6}
Zeal                             & CE                             & \multicolumn{1}{c|}{\cmark} & \multicolumn{1}{c|}{\xmark} & \multicolumn{1}{c|}{\cmark} & \cmark        & \multicolumn{1}{c|}{\cmark} & \multicolumn{1}{c|}{\cmark} & \multicolumn{1}{c|}{\cmark} & \cmark        & \cmark                                                                & 2                                                                                   & 3                                                                              \\ \hline
\rowcolor[HTML]{D0D0D0}
Coin98                           & CE                             & \multicolumn{1}{c|}{\cmark} & \multicolumn{1}{c|}{\cmark} & \multicolumn{1}{c|}{\cmark} & \notcheckmark & \multicolumn{1}{c|}{\cmark} & \multicolumn{1}{c|}{\xmark} & \multicolumn{1}{c|}{\cmark} & \notcheckmark & \cmark                                                                & 2                                                                                   & 2                                                                              \\ \hline
\rowcolor[HTML]{D0D0D0}
Exodus                           & CE                             & \multicolumn{1}{c|}{\cmark} & \multicolumn{1}{c|}{\cmark} & \multicolumn{1}{c|}{\cmark} & \xmark        & \multicolumn{1}{c|}{\cmark} & \multicolumn{1}{c|}{\xmark} & \multicolumn{1}{c|}{\cmark} & \xmark        & \cmark                                                                & 2                                                                                   & 2                                                                              \\ \hline
\rowcolor[HTML]{D0D0D0}
OKX                              & CE                             & \multicolumn{1}{c|}{\cmark} & \multicolumn{1}{c|}{\cmark} & \multicolumn{1}{c|}{\cmark} & \notcheckmark & \multicolumn{1}{c|}{\cmark} & \multicolumn{1}{c|}{\cmark} & \multicolumn{1}{c|}{\cmark} & \notcheckmark & \cmark                                                                & 2                                                                                   & 2                                                                              \\ \hline
\rowcolor[HTML]{D0D0D0}
Rabby                            & CE                             & \multicolumn{1}{c|}{\cmark} & \multicolumn{1}{c|}{\xmark} & \multicolumn{1}{c|}{\cmark} & \notcheckmark & \multicolumn{1}{c|}{\cmark} & \multicolumn{1}{c|}{\cmark} & \multicolumn{1}{c|}{\cmark} & \notcheckmark & \cmark                                                                & 2                                                                                   & 2                                                                              \\ \hline
\rowcolor[HTML]{D0D0D0}
Rainbow                          & CE                             & \multicolumn{1}{c|}{\cmark} & \multicolumn{1}{c|}{\cmark} & \multicolumn{1}{c|}{\cmark} & \xmark        & \multicolumn{1}{c|}{\cmark} & \multicolumn{1}{c|}{\xmark} & \multicolumn{1}{c|}{\cmark} & \xmark        & \cmark                                                                & 2                                                                                   & 2                                                                              \\ \hline
\rowcolor[HTML]{D0D0D0}
Sender                           & CE                             & \multicolumn{1}{c|}{\cmark} & \multicolumn{1}{c|}{\cmark} & \multicolumn{1}{c|}{\cmark} & \notcheckmark & \multicolumn{1}{c|}{\cmark} & \multicolumn{1}{c|}{\cmark} & \multicolumn{1}{c|}{\cmark} & \xmark        & \cmark                                                                & 2                                                                                   & 2                                                                              \\ \hline
\rowcolor[HTML]{D0D0D0}
Trust                            & CE                             & \multicolumn{1}{c|}{\cmark} & \multicolumn{1}{c|}{\cmark} & \multicolumn{1}{c|}{\cmark} & \xmark        & \multicolumn{1}{c|}{\cmark} & \multicolumn{1}{c|}{\xmark} & \multicolumn{1}{c|}{\cmark} & \xmark        & \cmark                                                                & 2                                                                                   & 2                                                                              \\ \hline
\rowcolor[HTML]{D0D0D0}
TokenPocket                      & CE                             & \multicolumn{1}{c|}{\cmark} & \multicolumn{1}{c|}{\cmark} & \multicolumn{1}{c|}{\cmark} & \xmark        & \multicolumn{1}{c|}{\cmark} & \multicolumn{1}{c|}{\xmark} & \multicolumn{1}{c|}{\cmark} & \xmark        & \cmark                                                                & 2                                                                                   & 2                                                                              \\ \hline
\rowcolor[HTML]{D0D0D0}
Tomato    & CE & \multicolumn{1}{c|}{\cmark} & \multicolumn{1}{c|}{\cmark} & \multicolumn{1}{c|}{\cmark} & \xmark & \multicolumn{1}{c|}{\cmark} & \multicolumn{1}{c|}{\cmark} & \multicolumn{1}{c|}{\cmark} & \xmark & \cmark & 2 & 2 \\ \hline
\rowcolor[HTML]{D0D0D0}
Wombat    & CE & \multicolumn{1}{c|}{\cmark} & \multicolumn{1}{c|}{\xmark} & \multicolumn{1}{c|}{\cmark} & \xmark & \multicolumn{1}{c|}{\cmark} & \multicolumn{1}{c|}{\cmark} & \multicolumn{1}{c|}{\cmark} & \xmark & \cmark & 2 & 2 \\ \hline
\rowcolor[HTML]{D0D0D0}
Zerion                           & CE                             & \multicolumn{1}{c|}{\cmark} & \multicolumn{1}{c|}{\cmark} & \multicolumn{1}{c|}{\cmark} & \xmark        & \multicolumn{1}{c|}{\cmark} & \multicolumn{1}{c|}{\xmark} & \multicolumn{1}{c|}{\cmark} & \xmark        & \cmark                                                                & 2                                                                                   & 2                                                                              \\ \hline
\rowcolor[HTML]{D0D0D0}
Enkrypt                          & CE                             & \multicolumn{1}{c|}{\cmark} & \multicolumn{1}{c|}{\cmark} & \multicolumn{1}{c|}{\cmark} & \xmark        & \multicolumn{1}{c|}{\xmark} & \multicolumn{1}{c|}{\xmark} & \multicolumn{1}{c|}{\xmark} & \xmark        & \cmark                                                                & 1                                                                                   & 2                                                                              \\ \hline
\rowcolor[HTML]{D0D0D0}
FoxWallet & CE & \multicolumn{1}{c|}{\xmark} & \multicolumn{1}{c|}{\cmark} & \multicolumn{1}{c|}{\cmark} & \xmark & \multicolumn{1}{c|}{\xmark} & \multicolumn{1}{c|}{\xmark} & \multicolumn{1}{c|}{\xmark} & \xmark & \cmark & 1 & 2 \\ \hline
\rowcolor[HTML]{D0D0D0}
Klever                           & CE                             & \multicolumn{1}{c|}{\cmark} & \multicolumn{1}{c|}{\cmark} & \multicolumn{1}{c|}{\cmark} & \xmark        & \multicolumn{1}{c|}{\xmark} & \multicolumn{1}{c|}{\xmark} & \multicolumn{1}{c|}{\xmark} & \xmark        & \cmark                                                                & 1                                                                                   & 2                                                                              \\ \hline
\rowcolor[HTML]{D0D0D0}
Pontem                           & CE                             & \multicolumn{1}{c|}{\cmark} & \multicolumn{1}{c|}{\cmark} & \multicolumn{1}{c|}{\cmark} & \xmark        & \multicolumn{1}{c|}{\xmark} & \multicolumn{1}{c|}{\xmark} & \multicolumn{1}{c|}{\xmark} & \xmark        & \cmark                                                                & 1                                                                                   & 2                                                                              \\ \hline
\rowcolor[HTML]{EDEDED}
Bitget                           & CE                             & \multicolumn{1}{c|}{\cmark} & \multicolumn{1}{c|}{\xmark} & \multicolumn{1}{c|}{\cmark} & \xmark        & \multicolumn{1}{c|}{\cmark} & \multicolumn{1}{c|}{\xmark} & \multicolumn{1}{c|}{\cmark} & \xmark        & \cmark                                                                & 2                                                                                   & 1                                                                              \\ \hline
\rowcolor[HTML]{EDEDED}
Coinbase                         & CE                             & \multicolumn{1}{c|}{\cmark} & \multicolumn{1}{c|}{\xmark} & \multicolumn{1}{c|}{\cmark} & \xmark        & \multicolumn{1}{c|}{\cmark} & \multicolumn{1}{c|}{\cmark} & \multicolumn{1}{c|}{\xmark} & \xmark        & \cmark                                                                & 2                                                                                   & 1                                                                              \\ \hline
\rowcolor[HTML]{EDEDED}
Infinity                         & Desktop                            & \multicolumn{1}{c|}{\cmark} & \multicolumn{1}{c|}{\xmark} & \multicolumn{1}{c|}{\cmark} & \xmark        & \multicolumn{1}{c|}{\cmark} & \multicolumn{1}{c|}{\xmark} & \multicolumn{1}{c|}{\cmark} & \xmark        & \cmark                                                                & 2                                                                                   & 1                                                                              \\ \hline
\rowcolor[HTML]{EDEDED}
\rowcolor[HTML]{EDEDED}
SafePal                          & CE                             & \multicolumn{1}{c|}{\cmark} & \multicolumn{1}{c|}{\xmark} & \multicolumn{1}{c|}{\cmark} & \xmark        & \multicolumn{1}{c|}{\cmark} & \multicolumn{1}{c|}{\xmark} & \multicolumn{1}{c|}{\cmark} & \xmark        & \cmark                                                                & 2                                                                                   & 1                                                                              \\ \hline
\rowcolor[HTML]{EDEDED}
WigWam                           & CE                             & \multicolumn{1}{c|}{\cmark} & \multicolumn{1}{c|}{\xmark} & \multicolumn{1}{c|}{\cmark} & \xmark        & \multicolumn{1}{c|}{\cmark} & \multicolumn{1}{c|}{\xmark} & \multicolumn{1}{c|}{\cmark} & \xmark        & \cmark                                                                & 2                                                                                   & 1                                                                              \\ \hline

\end{tabular}
}
\end{table*}
\subsection{Display of Phishing Transfers}
When condition \textbf{C1} is satisfied, we assess the crypto wallets' security using condition \textbf{C2}: display of phishing transfers, and condition \textbf{C3}: addresses are shortened. To quantify the risk, we define the following five risk levels and assign one of them to each tested wallet if the associated requirement is satisfied. The higher the level is, the more risky the wallet is.
\begin{itemize}
\item \textbf{Level 4}: display all phishing transfers;
\item \textbf{Level 3}: display fake-ETH or fake-USDT transfers;
\item \textbf{Level 2}: display zero-ETH or zero-USDT, or conditionally display fake transfers\footnote{for example, displaying the fake transfer as a contract interaction or only displaying the special fake transfer of 0 amount.};
\item \textbf{Level 1}: display dust-ETH or dust-USDT transfers;
\item \textbf{Level 0}: display none of the phishing transfers;
\end{itemize}
In the above definition, we consider level 4 extremely risky because all three types of phishing transfers are displayed, which thus poses the highest risk to the wallet user. We consider level 3 highly risky because, compared to zero-value and dust-value transfers, fake transfers can be crafted at a lower cost while looking highly similar to the legitimate transfer, hence posing a higher risk to the wallet user. Level 2 is considered less risky since zero-value transfers and conditionally displayed fake transfers will not look highly similar to legitimate transfers. For level 1, we consider displaying dust-value transfers less risky than level 2 because dust-value transfers incur a higher attack cost than zero-value transfers.

\textbf{Findings:} Since the 17 wallets in Table~\ref{tab:wallet_results_zero} do not satisfy \textbf{C1} and display no transfers, their risk level is thus 0. For the remaining 36 wallets listed in Table~\ref{tab:wallet_results}, they all shorten users' addresses. Among them, 6 wallets, such as Bybit, Core, Ctrl, and Frontier, display all three types of phishing transfers, which hence have the highest risk level (level 4: extremely risky). In addition, there are 10 wallets, such as Backpack, Gate, Hana, OneKey, Phantom, Taho, Uniswap, and Zeal, displaying either fake-ETH or fake-USDT transfers, but not all three types of phishing transfers. Hence, their risk level is assigned to 3 (highly risky). Moreover, there are 15 wallets displaying zero-ETH or zero-USDT transfers but not fake transfers, such as Coin98, Enkrypt, Exodus, Rabby, and Trust etc. Hence, their risk level is lower, which is assigned to 2. Finally, 5 wallets are assigned at risk level 1 due to only displaying the dust-ETH or dust-USDT transfers, including Bitget, Coinbase, etc. In Fig.~\ref{fig:level4}, we show the screenshots of the displayed transfers on 4 extremely risky wallets: Bybit, Core, Ctrl, and Frontier. The screenshots of other wallets are deferred to Appendix~\ref{sec:appendix:screenshots} due to page limit.

\begin{figure}[!ht]
   \centering
   \includegraphics[width=.5\textwidth]{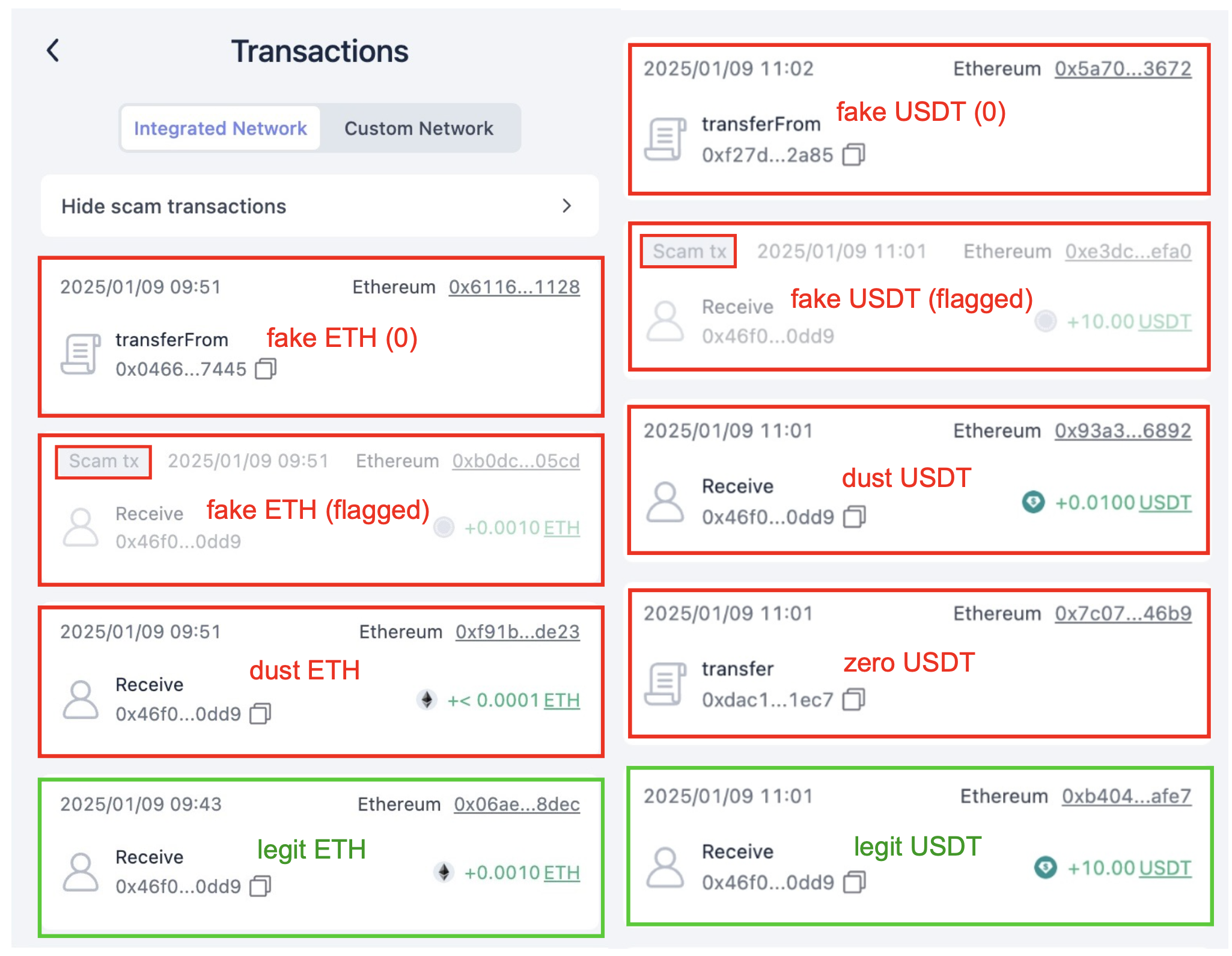}
   \caption{Transfers displayed on Rabby}%
   \label{fig:rabby}
\end{figure}

\textbf{Notable effort by Rabby:} In our results, it is worth mentioning that Rabby is the only one who flags phishing transfers. As shown in Fig.~\ref{fig:rabby}, both the fake ETH and USDT transfers are flagged as scam transactions. We hence give Rabby some credit for making such an effort in flagging phishing transfers sent by address poisoning attackers. However, our evaluation still suggests that more efforts are needed by Rabby, as dust-value transfers and the special fake transfers with 0 amount are not flagged yet.

\begin{table*}[!htbp]
\caption{Transaction activity provider of 17 wallets that do not display transfers (usability level = 0).}
\label{tab:provider0}
\centering
\begin{tabular}{|c|c|c|c|}
\hline
\textbf{Wallet} & \textbf{Entity} & \textbf{Domain Name} & \textbf{HTTP Response} \\ \hline
1chainAi           & Unknown            & https://api.chatgm.com    &         404 (Not Found)               \\ \hline
Aurox           & Self            & https://api.blockchain.getaurox.com    &    403 (Forbidden)             \\ \hline

Crypto.com      & Allnodes        & https://ethereum-rpc.publicnode.com    &      200 (Empty body)                 \\ \hline

Fluvi      & Ankr        & https://rpc.ankr.com/   &     200 (Invalid API key)           \\ \hline
Keplr           & Self            & https://evm-1.keplr.app                &      200 (Empty body)                  \\ \hline
MetaMask           & Self            & https://accounts.api.cx.metamask.io                &      200 (Empty body)                  \\ \hline
Nabox    & Publicnode        & https://ethereum.publicnode.com/   &   200 (Empty body)                     \\ \hline
Nightly      & Unknown        & https://mainnet.storyrpc.io/      &          200 (Empty body)               \\ \hline
Nufinetes      & Self        & https://api-common.nufinetes.com   &       200 (Empty body)                 \\ \hline

SubWallet       & Unknown         & https://mainnet.storyrpc.io            &   200 (Empty body)                \\ \hline
Virgo           & Base            & https://mainnet.base.org               &   403 (Forbidden)              \\ \hline
Zapit      & Self        & https://gateway.zapit.io               &       200 (Empty body)              \\ \hline
\begin{tabular}[c]{@{}c@{}}Arcana, Cypher, Frame, \\ Quantum,  JustLiquidity\end{tabular} & \multicolumn{3}{c|}{N/A} \\ \hline
\end{tabular}%
\end{table*}

\begin{table*}[!htbp]
\caption{Transaction activity provider of 36 wallets that display transfers (usability level > 0). "ZE", "FE" respectively stand for "zero-ETH", "fake-ETH" transfers. "ZU", "DU", "FU" respectively stand for "zero-USDT", "dust-USDT", "fake-USDT" transfers. "LU" stands for "legit-USDT" transfer.}
\label{tab:provider1}
\centering
\resizebox{\textwidth}{!}{%
\begin{tabular}{|c|cc|c|c|}
\hline
\multirow{2}{*}{\textbf{Wallet}} & \multicolumn{2}{c|}{\textbf{Transaction Activity Provider}}           & \multirow{2}{*}{\textbf{Filtered by Provider}} & \multirow{2}{*}{\textbf{Filtered by Wallet}} \\ \cline{2-3}
                                 & \multicolumn{1}{c|}{\textbf{Entity}} & \textbf{Domain Name}                   &                                           &                                                \\ \hline
\rowcolor[HTML]{A9A7A7}
Bybit                            & \multicolumn{1}{c|}{Self}            & https://api2.bybit.com                 & -                                         & -                                              \\ \hline
\rowcolor[HTML]{A9A7A7}
Core                             & \multicolumn{1}{c|}{Self}            & https://proxy-api.avax.network         & -                                         & -                                              \\ \hline
\rowcolor[HTML]{A9A7A7}
Ctrl                             & \multicolumn{1}{c|}{Self}            & https://gql-router.xdefi.services      & -                                         & -                                              \\ \hline
\rowcolor[HTML]{A9A7A7}

Safnect                             & \multicolumn{1}{c|}{Self}            & https://server.safnect.com      & -                                         & -                                              \\ \hline
\rowcolor[HTML]{A9A7A7}
Superhero                             & \multicolumn{1}{c|}{Etherscan}            & https://api.etherscan.io      & -                                         & -                                              \\ \hline
\rowcolor[HTML]{B6B6B6}
Backpack                         & \multicolumn{1}{c|}{Self}            & https://backpack-api.xnfts.dev         & ZE, ZU                                    & -                                              \\ \hline
\rowcolor[HTML]{B6B6B6}
Gate                             & \multicolumn{1}{c|}{Self}            & https://dapp.gateio.services           & ZE, ZU                                    & -                                              \\ \hline
\rowcolor[HTML]{B6B6B6}
Hana                             & \multicolumn{1}{c|}{Self}            & https://api.hanawallet.io              & ZE                                        & -                                              \\ \hline
\rowcolor[HTML]{B6B6B6}
Onekey                           & \multicolumn{1}{c|}{Self}            & https://wallet.onekeycn.com            & FU                                        & -                                              \\ \hline
\rowcolor[HTML]{B6B6B6}
Nest                             & \multicolumn{1}{c|}{Self}            & https://api.nestwallet.app      & -                                         & FE, ZU                                              \\ \hline
\rowcolor[HTML]{B6B6B6}
Nufi                             & \multicolumn{1}{c|}{Self}            & https://alchemy-evm.nu.fi      & -                                         & ZE, ZU                                              \\ \hline
\rowcolor[HTML]{B6B6B6}
Phantom                          & \multicolumn{1}{c|}{Self}            & https://api.phantom.app                & FU                                        & -                                              \\ \hline
\rowcolor[HTML]{B6B6B6}
Taho                             & \multicolumn{1}{c|}{Alchemy}         & https://eth-mainnet.alchemyapi.io      & ZU                                        & -                                              \\ \hline
\rowcolor[HTML]{B6B6B6}
Uniswap                          & \multicolumn{1}{c|}{Self}            & https://gateway.uniswap.org            & ZU                                        & -                                              \\ \hline
\rowcolor[HTML]{B6B6B6}
Zeal                             & \multicolumn{1}{c|}{Amazon AWS}      & https://eu-west-1.amazonaws.com        & ZE                                        & -                                              \\ \hline
\rowcolor[HTML]{D0D0D0}
Coin98                           & \multicolumn{1}{c|}{Etherscan}       & https://api.etherscan.io               & -                                         & FE, FU, ZU                                             \\ \hline
\rowcolor[HTML]{D0D0D0}
Exodus                           & \multicolumn{1}{c|}{Self}            & https://avax-c.a.exodus.io             & FE, ZU, FU                                & -                                              \\ \hline
\rowcolor[HTML]{D0D0D0}
OKX                              & \multicolumn{1}{c|}{Self}            & https://wallet.okex.org                &                                           & FE, FU                                               \\ \hline
\rowcolor[HTML]{D0D0D0}
Rabby                            & \multicolumn{1}{c|}{Self}            & https://api.rabby.io                   & ZE                                        & FE, FU                                              \\ \hline
\rowcolor[HTML]{D0D0D0}

Sender                           & \multicolumn{1}{c|}{Self}            & https://api.nearblocks.io              & FU                                        & FE                                              \\ \hline
\rowcolor[HTML]{D0D0D0}
Trust                            & \multicolumn{1}{c|}{Unknown}         & https://ethereum.twnodes.com           & FE, ZU, FU                                & -                                              \\ \hline
\rowcolor[HTML]{D0D0D0}
TokenPocket                      & \multicolumn{1}{c|}{Self}            & https://pretxs.mytokenpocket.vip       & -                                         & FE, FU, ZU                                     \\ \hline
\rowcolor[HTML]{D0D0D0}
Wombat                             & \multicolumn{1}{c|}{Etherscan}            & https://api.etherscan.io      & FE, FU                                         & ZE                                              \\ \hline
\rowcolor[HTML]{D0D0D0}
Enkrypt                          & \multicolumn{1}{c|}{Blockscout}      & https://eth.blockscout.com             & FE, LU, ZU, DU, FU                        & -                                              \\ \hline
\rowcolor[HTML]{D0D0D0}
Klever                           & \multicolumn{1}{c|}{Self}            & https://apis.klever.io                 & FE, LU, ZU, DU, FU                        & -                                              \\ \hline
\rowcolor[HTML]{D0D0D0}
Pontem                           & \multicolumn{1}{c|}{Self}            & https://control.pontem.network         & FE, LU, ZU, DU, FU                        & -                                              \\ \hline
\rowcolor[HTML]{EDEDED}
Bitget                           & \multicolumn{1}{c|}{Unknown}         & https://api-app-fast.chainnear.com     & ZE, FE, ZU, FU                            & -                                              \\ \hline
\rowcolor[HTML]{EDEDED}
Coinbase                         & \multicolumn{1}{c|}{Self}            & https://blockchain.wallet.coinbase.com & ZE, DU                                    & FE, FU                                         \\ \hline
\rowcolor[HTML]{EDEDED}
Safepal                          & \multicolumn{1}{c|}{Self}            & https://ap.isafepal.com                & FE, ZU, FU                                & ZE                                             \\ \hline
\rowcolor[HTML]{EDEDED}
Wigwam                           & \multicolumn{1}{c|}{Infura}          & https://mainnet.infura.io              & ZE, FE, ZU, FU                            & -                                              \\ \hline
\begin{tabular}[c]{@{}c@{}}Frontier, Rainbow, \\ Tomato, FoxWallet, \\ Infinity, Zerion\end{tabular} & \multicolumn{4}{c|}{N/A} \\ \hline
\end{tabular}%
}
\end{table*}

\subsection{Transaction Activity Provider}
Since both the wallets' usability and security are determined by the transfers displayed in the transaction activity entry, it is important to understand where and how the wallets obtain the users' transaction history. In fact, all crypto wallets rely on a backend service, which is synchronized with the Ethereum blockchain, to feed the users' transaction history. In this section, we choose to measure the backend services utilized by each crypto wallet. To achieve the goal, we leverage Chrome's inspection tool and Wireshark~\cite{me:wireshark} to capture the network traffic of each wallet and analyze the HTTP requests and responses. After that, we manually inspect each HTTP response to identify the backend URL that provides wallets with the transaction history. After identifying the backend URL queried by the wallet for downloading the transaction history, we then query whois~\cite{me:whois} to determine the entity that owns the domain name.

\textbf{Entity of provider:} In Table~\ref{tab:provider0} and \ref{tab:provider1}, we summarize the entity and domain name of the transaction activity providers used by the wallets. In addition, if the domain name is registered by the same entity operating the crypto wallet, the entity is presented as "Self". Otherwise, the entity is presented as either the third-party provider's name or "Unknown", depending on whether the registrar information is linkable to a known organization. Table~\ref{tab:provider0} and ~\ref{tab:provider1} respectively show the transaction activity providers of 12 wallets at usability level 0 and 30 wallets at usability level > 0. We are unable to identify the provider's URL of 5 wallets in Table~\ref{tab:provider0} and 6 wallets in Table~\ref{tab:provider1} due to that they use the TLS communication protocol, including Arcana, Cypher, Frame, Frontier, Rainbow, FoxWallet, etc. Nevertheless, from both tables, we can see that the majority of the crypto wallets are running their own backend services to feed users' transaction history. 11 wallets are using third-party providers, such as Crypto.com, Fluvi, Nabox, Virgo, Superhero, Taho, Zeal, Coin98, Wombat, Enkrypt, and Wigwam. There is a total of 9 unique third-party providers, including Allnodes, Ankr, Publicnode, Base, Etherscan, Alchemy, AWS, Blockscount, and Infura. Among them, the most popular provider is Etherscan, which serves 3 wallets. There are another 5 wallets using a transaction activity provider from an unknown/unlinkable entity, such as 1chainAi, Nightly, SubWallet, Trust, and Bitget.

\textbf{Impact on usability:} For the 12 wallets in Table~\ref{tab:provider0} at usability level 0, we examine the reason why they do not display any transfers. Specifically, we want to investigate whether the problem is caused by the transaction activity provider or the wallet itself. To answer this question, we manually analyze the payload of the captured HTTP packets between the wallet and the transaction activity provider. The HTTP response of each provider is summarized in Table~\ref{tab:provider0}. We can see that 9 wallets receive an HTTP response from their provider with a status code of 200. However, 8 of them receive empty data in the HTTP body. Ankr, the provider of the wallet Fluvi, is the only one that indicates errors in the wallet's HTTP request: "invalid API key". For the other 3 wallets, 1chainAi receives a reply of 404 from its provider, and Aurox and Virgo receive a reply of 403 from their provider. Overall, our analysis indicates that miscommunication with the transaction provider is the culprit for causing the 12 wallets not to display any transfers. To successfully retrieve users' transaction history, the wallets should either fix the errors in their requests or ask their provider to fix the errors in the response.

\textbf{Impact on security:} For the 30 wallets in Table~\ref{tab:provider1} at usability level > 0, five wallets display all three types of phishing transfers (risk level = 4), which means that their transaction activity providers do not filter out any phishing transfers. For the other 25 wallets that hide one or more phishing transfers (risk level < 4), it is interesting to see whether the phishing transfers are filtered out by the transaction activity provider or the wallet itself. Likewise, we answer this question by manually analyzing the payload of the captured HTTP packets between the wallet and the transaction activity provider. Our analysis result is summarized in Table~\ref{tab:provider1}. It can be seen that for the 10 wallets at risk level 3 (from Backpack to Zeal), 8 of them have the phishing transfers filtered out by their provider, 2 of them have the phishing transfers filtered out by the wallet itself. Among them, 7 wallets filter out only zero-value transfers (zero-ETH and zero-USDT), and three wallets filter out only one of the two fake transfers (fake-USDT and fake-ETH). Besides, for the 11 wallets at risk level 2 (from Coin98 to Pontem), 3 wallets filter out phishing transfers solely on the wallet end, 5 wallets filter out phishing transfers solely on the provider end, and 3 wallets filter them out collaboratively on both the provider and wallet ends. Among them, the providers of wallet Enkrypt, Klever, and Pontem have the strongest impact for filtering out all kinds of USDT transfers, including the legitimate ones. For the last four wallets at risk level 1 (from Bitget to Wigwam), two wallets have the zero-value transfers and fake transfers filtered out solely by their provider, and the other two wallets filter out zero-value and fake transfers jointly on the provider and wallet ends. Overall, our analysis indicates that most wallets rely on their transaction provider to detect and filter out phishing transfers. Some wallets choose to detect phishing transfers themselves directly on the wallet end. In contrast, only a small fraction of wallets detect phishing transfers jointly on the provider and wallet ends. However, the capability of detecting phishing transfers varies by provider and wallet. Among them, the providers of wallet Bitget and Wigwam have the strongest detection capability, and wallet TokenPocket has the strongest detection capability.

\textbf{Misuse of Etherscan's API:} It is also worth mentioning that wallet Superhero, Coin98, and Wombat all use Etherscan as the transaction activity provider. However, Superhero displays all three types of phishing transfers, while Coin98 and Wombat have fake transfers filtered out by Etherscan. Our investigation indicates that this is because Coin98 and Wombat both specify the legitimate USDT token's address in the request (e.g., through a parameter of "\&contract\_address"), but Superhero does not, which thus causes Etherscan to return both fake-ETH and fake-USDT transfers.

\begin{figure*}[!htbp]
  \centering
    \subfloat[Alert of high-risky address (OKX)]{%
  \includegraphics[width=0.3\textwidth]{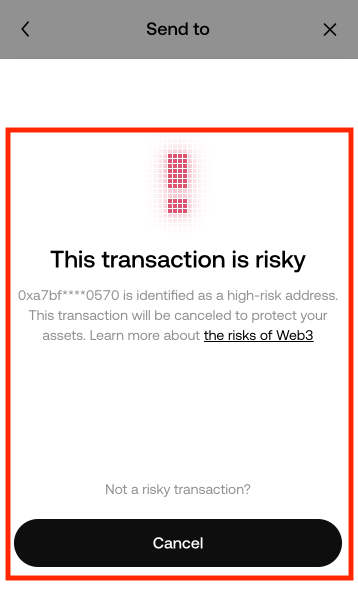}
  \label{fig:riskyalert}}%
   \subfloat[Confirmation required when transferring to an unknown address (Rabby)]{%
  \includegraphics[width=0.338\textwidth]{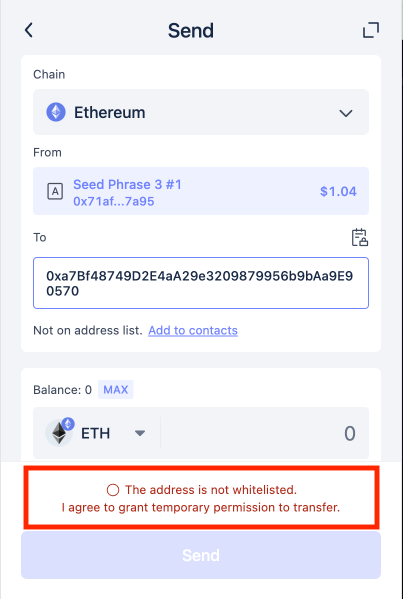}
  \label{fig:popupreminder}}
   \subfloat[Reminder of tranferring to an unknown address (Zeal)]{%
  \includegraphics[width=0.304\textwidth]{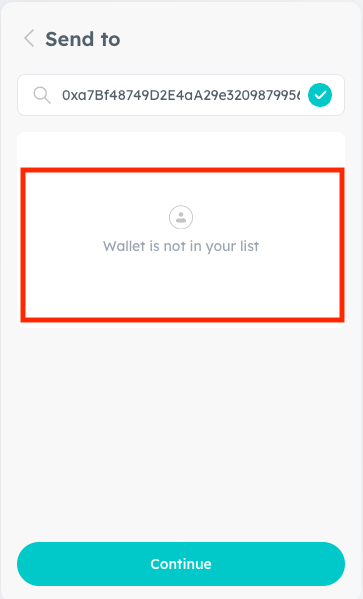}
  \label{fig:plainreminder}}
  \caption{Three types of transaction warning features when transferring to a phishing address.}%
  \label{fig:warning}
\end{figure*}
\subsection{Transaction Warning Features}
Since the ultimate goal of the address poisoning attack is to lure victims into transferring funds to the phishing address, it is thus crucial for the wallets to prevent users from interacting with phishing addresses involved in address poisoning or at least warn users about the potential risk. To evaluate whether crypto wallets have employed such a preventive or warning feature, we test each wallet by initiating transactions to transfer funds to the phishing address. Specifically, on each wallet, we initiate transactions to attempt to transfer both ETH and USDT from the victim address \textbf{V} to the phishing address \textbf{A} that we generated in the previous experiment. Since address \textbf{A} (\hash{0x46f0042749ad2383471639b57833cd80bf1f0dd9}) has not been flagged on Etherscan, for comparison, we also initiate transactions to transfer funds to the notorious phishing address \textbf{F} (\hash{0xa7Bf487***E90570}) that stolen 20 million USDT from Binance~\cite{bnblost} through the address poisoning attack, which has been flagged on Etherscan. Before signing the transaction on each wallet, we check whether there are warning messages displayed or preventive measures that alert us about the potential risk.

\begin{table}[]
\caption{Transaction Warning Results}
\label{tab:addr_warning}
\resizebox{\columnwidth}{!}{%
\begin{tabular}{|c|cccc|}
\hline
\multirow{2}{*}{\textbf{Wallet}}                                                                           & \multicolumn{2}{c|}{\textbf{ETH Recipient}}                                                        & \multicolumn{2}{c|}{\textbf{USDT Recipient}}                                  \\ \cline{2-5} 
                                                                                                  & \multicolumn{1}{c|}{\textbf{A}} & \multicolumn{1}{c|}{\textbf{F}} & \multicolumn{1}{c|}{\textbf{A}} & \textbf{F} \\ \hline
Bybit                                                                                             & \multicolumn{1}{c|}{-}                 & \multicolumn{1}{c|}{Alert}             & \multicolumn{1}{c|}{-}                 & Alert             \\ \hline
Ctrl                                                                                              & \multicolumn{1}{c|}{-}                 & \multicolumn{1}{c|}{Alert}             & \multicolumn{1}{c|}{-}                 & Alert             \\ \hline
OKX                                                                                               & \multicolumn{1}{c|}{Reminder}          & \multicolumn{1}{c|}{Alert}             & \multicolumn{1}{c|}{Reminder}          & Alert             \\ \hline
\begin{tabular}[c]{@{}c@{}}Onekey, Phantom, \\ Rabby, Uniswap \end{tabular} & \multicolumn{4}{c|}{Confirmation required} \\ \hline
\begin{tabular}[c]{@{}c@{}}Arcana, Nest, \\ Quantum, Zeal \end{tabular} & \multicolumn{4}{c|}{Reminder} \\ \hline
\end{tabular}%
}
\end{table}
Our testing results are summarized in Table \ref{tab:addr_warning}. Among the 53 tested wallets, we found that 11 wallets employed different transaction warning features. Specifically, when we intend to transfer to the flagged phishing address \textbf{F}, three wallets, Bybit, Ctrl, and OKX, display a clear message warning that our transaction is highly risky, as shown in Fig.~\ref{fig:riskyalert}. However, no warning is given when we transfer to phishing address \textbf{A}, except OKX, which reminds us that \textbf{A} is an unknown address. Besides, when we transfer to both \textbf{A} and \textbf{F} on the other four wallets, OneKey, Phantom, Rabby, and Uniswap, a transaction confirmation window is popped up, which indicates that the recipient address is unknown (e.g., not in the contact list, had no prior interactions). To close the window and proceed with the transfer, we need to click the confirmation button. In Fig.~\ref{fig:popupreminder}, we show such a transaction confirmation feature. In addition, there are another four wallets, Zeal, Arcana, Nest, and Quantum, that display a warning message showing the recipient \textbf{A} and \textbf{F} are unknown addresses (e.g., not in contact list), as can be seen in Fig.~\ref{fig:plainreminder}. However, no confirmation is required to proceed with the transfer. In comparison, all the remaining 42 wallets do not have any forms of transaction warning or preventive measures employed when we attempt to transfer to \textbf{A} and \textbf{F}.

In summary, our evaluation results suggest that transaction warning has not become a common feature in the design of Ethereum crypto wallets, as only three wallets, Bybit, Ctrl, and OKX, display a clear warning message when we attempt to transfer to the phishing address. This highlights a significant security gap in most crypto wallets regarding users' fund protection. While there are some wallets supporting transaction confirmation features, they simply remind us that the recipient address is unknown without explicitly informing us that we are interacting with phishing addresses. Finally, we give credit to the three wallets, Bybit, Ctrl, and OKX, that explicitly alert us about the phishing risk, which is a good preventive countermeasure. However, such an alert is only triggered when the recipient address has already been flagged. Hence, future improvements are needed for them to alert users about the risk of interacting with phishing addresses that have not been flagged.
 
\section{Discussion}
\label{sec:discuss}
Overall, our analysis suggests that among the 53 Ethereum crypto wallets, 17 wallets cannot successfully display the users' transaction history, most of which are caused by communication failures with their transaction activity provider. Among the 36 wallets that successfully display transaction history, 16 wallets cannot distinguish fake transfers from legitimate transfers, which thus pose a high risk to their users. While there are 15 wallets that hide fake transfers, they are still risky due to displaying zero-value phishing transfers. Only 5 wallets are less risky for displaying only the dust-value phishing transfers. Moreover, only three wallets throw a clear warning message to indicate the risk when users attempt to transfer funds to the phishing address. Therefore, our work implies that it is imperative for the broader crypto wallet developer community to mitigate such a problem. Below, we first discuss what we expect to be achieved in an ideal crypto wallet. Then, we discuss our bug disclosure process and our advice for the crypto wallet users.

\subsection{An Ideal Crypto Wallet}
Our evaluation shows that no crypto wallets have achieved the highest standard for both usability and security. Here, we discuss what we expect to be achieved by an ideal wallet.

At a high level, an ideal wallet should provide the highest usability and present its users with the lowest risk. Specifically, the ideal wallet should display all legitimate token transfers and transactions, including ETH, USDT, and other legitimate tokens, ensuring that users have access to all legitimate transaction activities. Meanwhile, the wallet should also hide or flag all phishing transfers and transactions to prevent address poisoning attacks from misleading users. To accurately detect the phishing transfers, the wallet can treat fake token transfers, zero-value transfers, and dust-value transfers as suspicious transfers and then match them with a previous legitimate transfer by comparing the address similarity. If the addresses look highly similar, then the suspicious transfers should be flagged. In addition, the ideal wallet should also employ fund recipient verification mechanisms. That is, when the user attempts to transfer assets to a recipient address, the wallet should check whether the address has been flagged by trustworthy sources such as Etherscan and other scam alerting services~\cite{me:chainabuse, me:scamsniffer, me:forta}. If so, the wallet should send an explicit warning message to prompt the user about the potential risk. Moreover, if the recipient address is not flagged by trustworthy sources but is involved in the phishing transfers detected from the user's transaction activity, the wallet should also flag the address as highly risky and alert the user before sending the transaction.

In summary, we believe that the ideal wallet should provide the highest usability and the lowest risk. Meanwhile, the wallet should deploy proactive fund recipient verification mechanisms, which is the last line of defense against the address poisoning attack. By achieving such goals, the wallet can provide users with a high quality service while also safeguarding their crypto assets.

\subsection{Bug Report}
We have reported the bugs and security risks to the Ethereum crypto wallet developer community. As of this writing, 11 wallets have replied to our bug reports, and 8 have acknowledged our reported problem and are currently deploying countermeasures. While we are awaiting the responses from other crypto wallets, we briefly discuss the mitigation plans of those who confirmed our bug reports. Specifically, Onekey plans to roll out an address book feature and remind users with a highlighted window when the funds are sent to an unknown address. They will also ask users to double-check the transaction details each time. Phantom, which currently displays a warning message when users transfer funds to an unknown address, plans to hide all three types of phishing transfers in the transaction activity feed. Enkrypt confirms that address poisoning is a possible vulnerability, however, they argue that the attack requires human errors and hence do not plan to deploy preventive countermeasures. 

\subsection{Advice for Wallet Users}
While crypto wallets play a crucial role in protecting users' crypto assets, to more effectively mitigate the threat of address poisoning attacks, we also recommend countermeasures for individual wallet users. The most important one is to choose a wallet that provides the strongest security countermeasures, such as phishing transaction detection and labeling, fund recipient address verification, etc. In addition, users should remain vigilant when checking their transaction history and take cautious actions when copying and pasting addresses. It is always a good strategy to verify each character and ensure that the copied address is owned by the correct recipient. Additionally, sending a small test transfer before making a large-value transfer can further confirm that the address is correct and funds will arrive as expected. Another countermeasure that users can adopt is to register an Ethereum Name Record (ENR) for their addresses and use them to send or receive funds, especially when dealing with large-value transfers or sending regular payments to the same recipient. By using the human-readable name instead of the hexadecimal address, users can avoid mistakenly copying phishing addresses that resemble legitimate ones.

\section{Related Work}
\label{sec:relate}
In the existing literature, various cryptocurrency scams and phishing attacks have been studied, including Ponzi~\cite{kell2021forsage, bian2021image, xia20covidscams, bartoletti2020dissecting, bartoletti2018data, chen2018detecting}, fake exchange scams~\cite{xia2020characterizing}, phishings~\cite{chen2020phishing, badawi2020automatic, he2023txphishscope}, giveaway scams~\cite{xia20covidscams, vakilinia2022cryptocurrency, xigao2023doublenothing, li2023understanding}, honeypot contract scams~\cite{torres2019art, chen2020honeypot}, scam tokens~\cite{gao2020tracking, xia21scams}, and token theft~\cite{chen2023understanding}. The most relevant work to ours is Guan and Li~\cite{guan24ccs}, Tsuchiya et al.~\cite {tsuchiya2025blockchainaddresspoisoning}, and Chen et al.~\cite{chen2025dissecting}, which have studied the three types of phishing transfers utilized in the token-based address poisoning attack. Specifically, Guan and Li~\cite{guan24ccs} developed a detection system and detected over 16 million phishing transfers and 6 million phishing addresses on the Ethereum mainnet. Their work showed that more than 1800 victim transactions lost nearly 100 million US dollars. The similar findings were also reported in Tsuchiya et al.~\cite{tsuchiya2025blockchainaddresspoisoning}. Chen et al.~\cite{chen2025dissecting} studied four transaction payload-based phishing scams on Ethereum, including ice phishing, NFT order, address poisoning, and payable function scams. Under the address poisoning category, their work reported that more than 1000 victim transactions lost over 60 million US dollars. Despite the extensive analysis of the address poisoning activities on the Ethereum blockchain, there is a lack of systematic analysis of Ethereum crypto wallets' usability and security under the address poisoning. To our best knowledge, our work is the first one that conducts such a systematic analysis.

\section{Conclusion}
\label{sec:conclude}
This paper systematically evaluates the usability and security of 53 Ethereum crypto wallets under the address poisoning attack. The evaluation result shows that no wallet achieves the highest usability and security standard, implying further efforts are needed by the broader crypto wallet developer community to address such a problem.

 
\bibliographystyle{plainnat}
\bibliography{shixuan, bkc, bkcscam, bkc2, bkcscam2, twitterscam, youtubescam}

\appendix
\section{Ethical Consideration}
In this work, we have taken cautious actions to design our experiments. First, our testing transactions are conducted among three Ethereum addresses that are under our control. Although our testing transactions are eventually included in the Ethereum mainnet, they do not affect other addresses or users on Ethereum. Besides, our analysis of 53 Ethereum crypto wallets is also conducted on our own laptop, which is a controlled environment and does not affect other wallet users. Moreover, throughout the paper, we have tried our best to anonymize phishing addresses controlled by others through shortening their addresses. Finally, we have reported the usability bugs and security risks to the crypto wallet developer community and recommended mitigation solutions against the address poisoning threat. We hope to help them achieve the highest usability and security standard.  

\section{More Screenshots of Ethereum Wallets}
\label{sec:appendix:screenshots}
\begin{figure}[!htbp]
  \centering
  \subfloat[Nest.]{%
    \includegraphics[width=0.48\columnwidth]{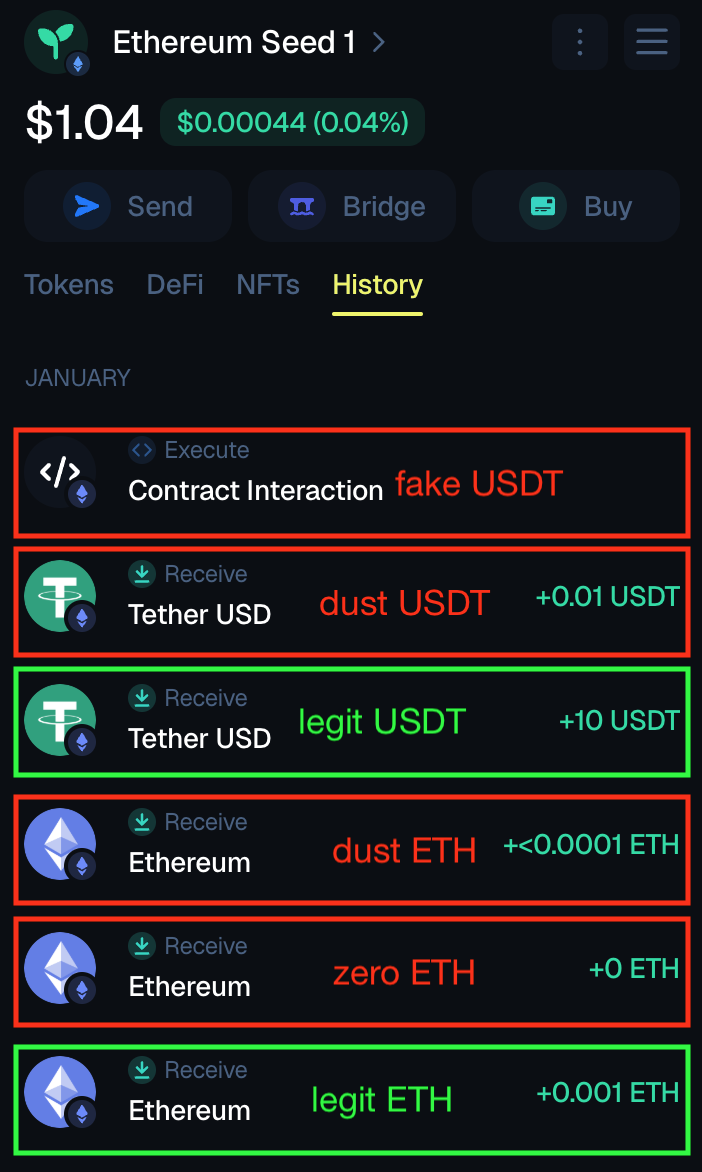}
    \label{fig:nest}
  }%
  \hfill
  \subfloat[Backpack.]{%
    \includegraphics[width=0.48\columnwidth]{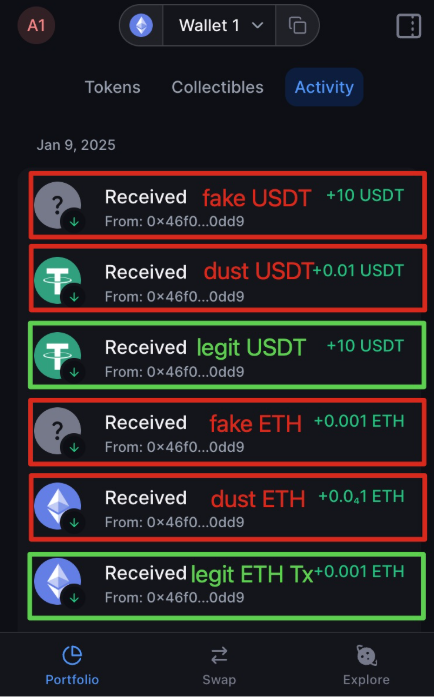}
    \label{fig:backpack}
  }%
  \caption{Screenshots of digital wallets at risk level 3.}
  \label{fig:wallet_risk3}
\end{figure}

\begin{figure}[!htbp]
  \centering
  \subfloat[Zerion.]{%
    \includegraphics[width=0.48\columnwidth]{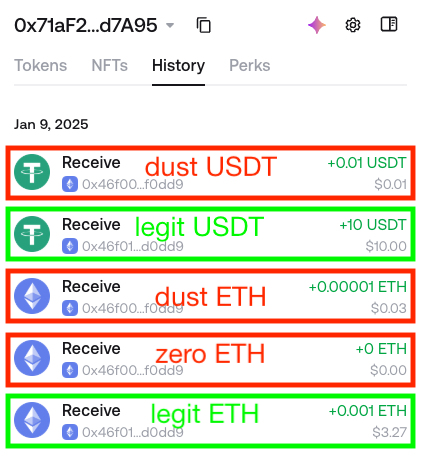}
    \label{fig:coinbase}
  }%
  \subfloat[Rainbow.]{%
    \includegraphics[width=0.48\columnwidth]{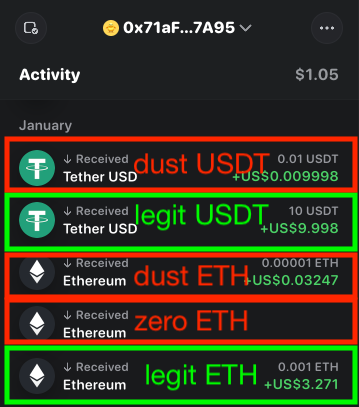}
    \label{fig:safepal}
  }%
  \caption{Screenshots of digital wallets at risk level 2.}
  \label{fig:wallet_risk2}
\end{figure}

\begin{figure}[!htbp]
  \centering
  \subfloat[Coinbase.]{%
    \includegraphics[width=0.48\columnwidth]{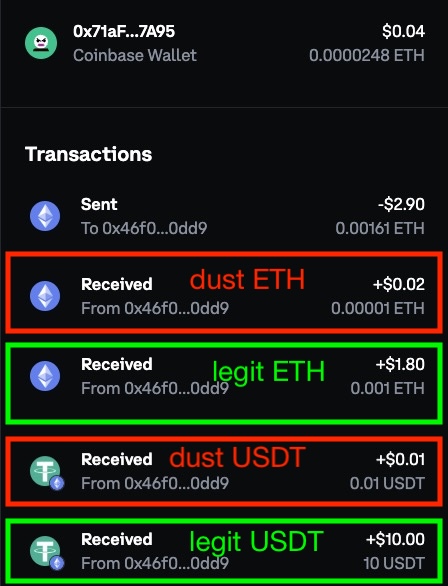}
    \label{fig:coinbase}
  }%
  \subfloat[Bitget.]{%
    \includegraphics[width=0.48\columnwidth]{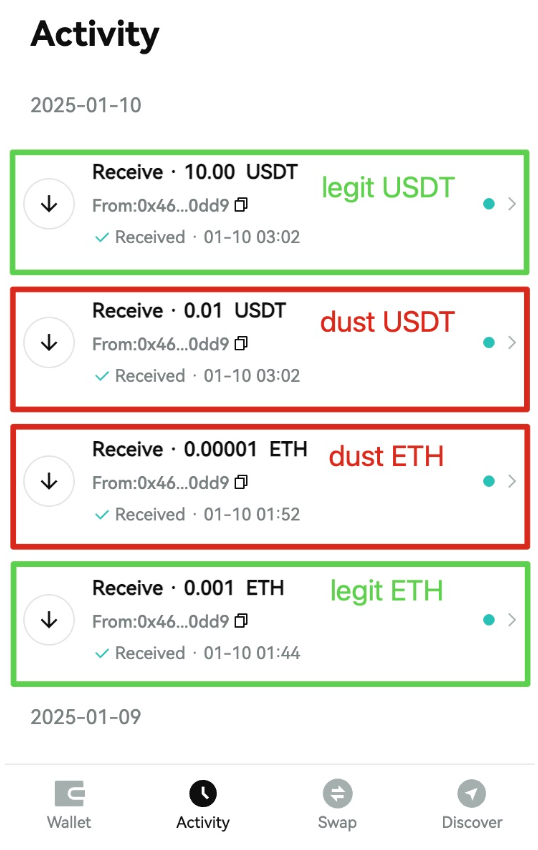}
    \label{fig:bitget}
  }%
  \caption{Screenshots of digital wallets at risk level 1.}
  \label{fig:wallet_risk1}
\end{figure}

This section presents the screenshots of displayed transfers on other crypto wallets. Fig.~\ref{fig:wallet_risk3} shows the transfers displayed on wallet Nest and Backpack, which were assigned at risk level 3. Fig.~\ref{fig:wallet_risk2} shows the transfers displayed on wallet Zerion and Rainbow, which were assigned at risk level 2. Fig.~\ref{fig:wallet_risk1} shows the transfers displayed on wallets Coinbase and Bitget, which were assigned at risk level 1.

\end{document}